United Nations University
Institute in Macau

www.unu.edu

UNU RESEARCH REPORT

# AI Systems as Digital Public Goods

EVIDENCE AND RECOMMENDATIONS
FROM A MULTI-STAKEHOLDER ASSESSMENT

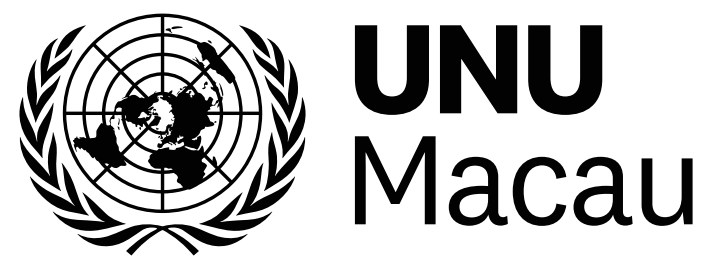

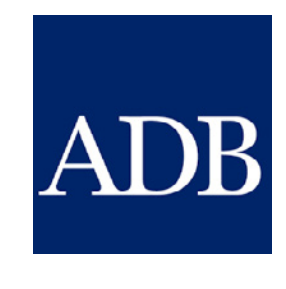

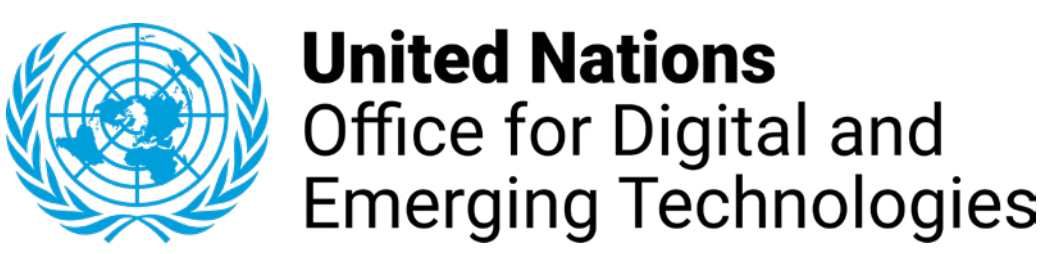

## Disclaimer

## Please direct inquiries to:

United Nations University Institute in Macau

**info_macau@unu.edu**

## Notes

In this publication, ADB recognizes "Macau" as Macau, China, and "China" as the People's Republic of China.

## Suggested citation:

## Table of Contents

## Table of Figures

## List of Tables

# Foreword

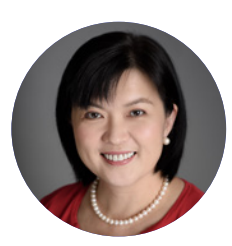

from **Stephanie KC Hung,**

Director General, Information Technology Department
Asian Development Bank

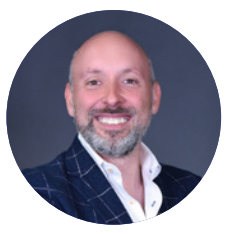

and **Antonio G. Zaballos,**

Director, Digital Sector Office
Asian Development Bank

Digital public goods (DPGs) are open-source software, data, standards and content that adhere to strong safeguards and accelerate progress toward the Sustainable Development Goals (SDGs). In 2024, the Asian Development Bank (ADB) joined the Digital Public Goods Alliance (DPGA), which issues DPG certifications. DPG certification is more than a label; it is a process designed to share best practices and enable replication across institutions. As a DPGA member, ADB is committed to submitting its digital solutions for the Alliance's certification to enable their wider reuse by governments and other stakeholders.

The year 2024 was pivotal in the development of AI, as the technology transitioned from the hype phase to practical implementation and adoption. As the leading multilateral financial institution in Asia and the Pacific, the ADB plays a critical role in ensuring that AI's transformative potential is harnessed not only to improve operational efficiency, but also to support inclusive, sustainable, and open development across the region. AI has the potential to offer unprecedented opportunities for developing member countries, including advancing digital public infrastructure, reducing costs, and improving essential public services such as health, education, and social protection. But what does it take for an AI system to qualify as a digital public good (AIDPG)? This report, drawing on perspectives from experts from developed and developing countries, seeks to answer that question.

We consulted stakeholders from academia, the United Nations system, the private sector, and civil society to share their insights on whether the current definition of DPGs is fit for purpose for AI and if refinements are potentially needed over time as the field continues to evolve. They also shared their thoughts on the challenges of recognizing AIDPGs, as well as the risks, benefits, and opportunities they pose.

These insights underscore the need for a nuanced approach to recognizing AIDPGs, given the inherent complexities of AI. For example, safeguarding standards typically required for DPGs—such as "privacy by design" and "do no harm" principles—must be systematically embedded to guarantee that AI, produced for the public good remains inclusive, trustworthy, and aligned with public interest. Furthermore, while open-source code is fundamental to DPGs, this approach may be reductive for AI, which also relies on components such as training data and model weights that go beyond code. Requiring full openness of these components may introduce risks of misuse and undermine public interest objectives, potentially conflicting with the principles DPGs are intended to uphold.

This work, aligned with ADB's Strategy 2030 and ADB Board Direction on Digital Transformation for Development, reflects strong collaboration with partners. We are especially grateful to the United Nations University in Macau, the United Nations Office of Digital and Emerging Technologies, and the Digital Public Goods Alliance (DPGA), as well as the wider expert community.

We hope this report will inform policy dialogue and support continued research and innovation in leveraging AIDPGs to address complex development challenges.

# Foreword

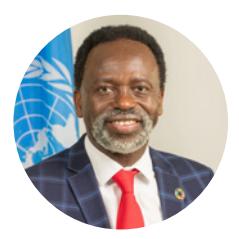

from **Professor Tshilidzi Marwala,**

Rector of the United Nations University and Under-Secretary-General of the United Nations

The most valuable creations are often those we can share without exhausting them. Knowledge is the clearest example: a good idea shared costs nothing, yet multiplies. Artificial intelligence seems to promise the same abundance: a model, once trained, can serve millions of people. But this promise masks a more complex problem. Releasing a system under an open-source license is not the same as making it genuinely accessible to those who need it most.

That distinction is precisely what this report sets out to examine. An AI system can be distributed under an open-source license and still remain out of reach for most because the data, computing infrastructure, and technical expertise required to operate it effectively are concentrated elsewhere. The term "open" describes the license of a system, while the expression "public good" describes who it truly benefits. The real work of development lies in closing the distance between the two.

As the academic arm of the UN, the United Nations University's mission is to bridge the gap between the academic world and the UN system, connecting research to public policy and fostering global knowledge exchange to ensure the fruits of scientific progress benefit humanity as a whole. This report is a direct expression of that mission. It examines what it would take for an AI system to be recognized as a digital public good, and it answers that question by going to the source: the people who design, fund, govern, and depend on these systems, in both developed and developing countries.

The findings invite careful reflection. Work is underway to adapt the digital public goods certification standard to include AI, and the UN Global Digital Compact has recognized open AI models as tools that can help societies address their own development priorities. Yet not a single AI system has been formally recognized as a digital public good under that standard. That is the gap this report confronts – the distance between what the international community has declared as a goal and what exists in practice to achieve it.

The report takes that gap seriously. It argues that openness is a matter of degree, not a binary condition, and that an AI model must be assessed on its own terms, not solely in the context of the broader system in which it operates. It also refutes the simplistic notion that what benefits the public in one country will similarly benefit the public in all countries. Shared resources do not manage themselves. Their value stems from the rules and institutions we put in place, and AI is no exception. Whether it becomes a common good or another advantage concentrated in the hands of the few will depend less on its capabilities than on how we choose to govern it.

This report was made possible through genuine collaboration. I thank the Asian Development Bank, whose financial support made this study possible, the United Nations Office for Digital and Emerging Technologies, The Digital Public Goods Alliance and the researchers and experts whose analysis and judgment informs every page. A digital public good is, above all, a commitment to future beneficiaries. We are not there yet, but this report lays out the conditions that would take us there.

# Foreword

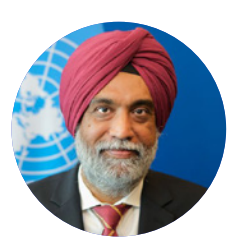

from **Amandeep Singh Gill**,

Under-Secretary-General and Special Envoy for Digital and Emerging Technologies at the United Nations Office for Digital and Emerging Technologies

When I had the privilege of supporting the United Nations Secretary-General's High-level Panel on Digital Cooperation in 2018, a key question before us was: how can digital technologies serve humanity as a whole? One of the answers that emerged was the concept of Digital Public Goods — open digital solutions that can help advance the Sustainable Development Goals and ensure that the benefits of technological innovation are shared widely rather than concentrated narrowly.

In the years since, I have witnessed how governments, international organizations, civil society, and industry have embraced this concept. What began as an idea aimed at fostering greater digital cooperation has evolved into a globally recognized framework for advancing inclusive and sustainable digital transformation. The Digital Public Goods Standard has become an important reference point for promoting openness, inclusion, and public value in the digital age.

Today, however, we find ourselves at another inflection point. Artificial Intelligence is challenging many of our existing assumptions. AI is transforming how knowledge is created, how services are delivered, and opportunities distributed. At the same time, it raises new questions about what openness, accessibility, accountability, and public benefit mean in practice. An AI model may be openly available, yet remain inaccessible to many because of the data, computing resources, and expertise required to use it effectively.

This report arrives at an important moment. It examines whether existing Digital Public Goods standards are sufficient for the age of AI and explores how they might evolve to reflect new technological realities. Weaving perspectives from policymakers, researchers and practitioners across developed and developing countries, it provides a thoughtful assessment of the challenges that AI presents to the philosophy and practice of public interest technology.

Its central message: the principles that underpin Digital Public Goods remain relevant as ever, but their application must adapt to ensure that AI serves the public interest. As AI becomes increasingly influential in shaping development, education, healthcare, scientific discovery, and economic opportunity, we must ensure that its benefits are accessible to all, equitable in terms of impact, and aligned with human well-being.

I welcome this report as a timely contribution to an essential global conversation. I am sure that its findings will provide valuable guidance to policymakers, standard-setting bodies, and industry leaders working to ensure that AI trends toward a global public good.

## About the authors

SERGE STINCKWICH

Dr. Serge Stinckwich is a computer scientist and a Senior Research Fellow at United Nations University Institute Office in Paris, with over 16 years of experience at the intersection of digital technologies and sustainable development across Asia and Africa. From March 2020 to March 2026, he was Head of Research and led an interdisciplinary research team at UNU Macau, focusing on responsible AI, gender and technology, digital health, and agent-based/participatory modelling to advance the Sustainable Development Goals (SDGs). His research spans complex systems modelling, social simulation, and the impact of AI on global challenges. Previously, he served as Associate Professor and researcher in an international joint research unit between the French Research Institute on Sustainable Development (IRD), Sorbonne University and five universities located in Cameroon, Morocco, Senegal and Vietnam. Passionate about leveraging technology for inclusive and sustainable futures, Dr. Stinckwich also conducts training sessions for policymakers and stakeholders on AI and climate, Large Language Models and synthetic data, and their implications for SDGs, as well as using agent-based models to design better health policies. Dr. Stinckwich has been an open-source advocate since 1992.

NATALIE WONG

Dr. Natalie Wong is a Visiting Research Fellow at the United Nations University Institute in Macau (UNU Macau), where her research sits at the intersection of artificial intelligence, human development, and technology policy. Holding a PhD in Psychology from the Chinese University of Hong Kong (CUHK), she has collaborated with UN agencies and Multilateral Development Banks to translate complex, cross-cultural data into actionable insights. As a mixed-methods expert, Dr. Wong's current work focuses on evaluating AI systems as a digital public good, exploring the impact of generative AI on ethics and education, and addressing systemic inequalities in digital ecologies. She is actively collaborating with UN Women on cybersecurity and women's leadership to develop targeted solutions for navigating digital transformations. Rooted in her training as a psychologist, Dr. Wong is deeply interested in how digital ecologies impact human development. She specializes in synthesizing voices from diverse backgrounds, ensuring that lived human experiences serve as the foundation for technology policy. Her research has been published in leading academic journals, as well as upcoming UN and ADB reports.

ALLY S. NYAMAWE

Dr. Ally S. Nyamawe is a computer scientist and academic with over 15 years of experience in research, teaching, and innovation. He is currently a researcher at the United Nations University Institute in Macau (UNU Macau). Prior to joining UNU Macau in 2024, he served at the University of Dodoma, Tanzania, in various academic ranks since 2009. Dr. Nyamawe leads and contributes to collaborative initiatives that promote the adoption of AI technologies in Tanzania. His research spans applied AI in software engineering and digital health, with a focus on digital health transformation in Africa. Dr. Nyamawe's work centres on bridging research, policy, and practice to support the responsible adoption of AI, ensuring its ethics, inclusivity, and alignment with national and global development priorities. Through his contributions to research, implementation, and policy advisory, he is committed to advancing resilient, human-centred AI systems that deliver meaningful societal impact.

JIA'AN LIU

Jia'an LIU is a computer scientist and research assistant at the United Nations University Institute in Macau. His research lies at the intersection of agentic AI, computational social science, and AI for sustainable development, with particular expertise in large language model-enabled agent-based modelling, multi-agent systems, urban and social simulation, and policy-oriented digital twin frameworks. His work develops spatially grounded, LLM-enhanced simulation systems to study complex societal challenges such as urban resilience, climate adaptation, healthcare accessibility, misinformation diffusion, inclusive education, and digital inclusion. His broader technical background spans machine learning, computer vision, remote sensing, and uncertainty-aware video motion analysis, alongside the design of AI workflows, intelligent knowledge pipelines, and research infrastructure for policy and scientific applications. His recent research also examines AI open-source readiness, and the evaluation of large language models for social policymaking, bridging technical innovation with questions of governance, openness, and public value.

FARHAN LATIF

Dr. Farhan Latif is a Research Fellow at UNU Macau. He is a transdisciplinary social scientist with expertise in the social and policy implications of digital technologies for sustainable development. His ongoing research projects at UNU Macau examine the implications of AI and digital platforms for the future of work, as well as how the political economy of digital technologies is shaping cyberspace and AI governance. He holds a PhD in Sustainability Management and a Master's in Development Practice from the University of Waterloo, Canada. His experience spans policy research with multilateral institutions and on-the-ground experience implementing technology-for-development initiatives in Kenya, Pakistan, and Canada. His research and policy perspectives are shaped by direct field engagement with the practical challenges of reducing digital divides and promoting more equitable development in traditionally marginalized settings.

JAIMEE STUART

Dr. Jaimee Stuart is Head of Research (Acting) at UNU Macau. She is an applied cultural and developmental psychologist specializing in cyberpsychology and the social impacts of emerging technologies. Her work spans computer-human interaction, social justice, peace, and development with a particular focus on the creation of safe and secure online spaces and the empowerment of minorities (cultural, religious, gender, and sexual orientation) and those who experience inflated risk factors (e.g. exposure to violence, low socio-economic status, displacement). Dr. Stuart leads flagship projects across gender, AI and cybersecurity as well as on youth perspectives and experience with AI emerging digital technologies. Before joining UNU Macau, Dr. Stuart served as Research and Evidence Lead at Pathways in Place, Griffith University, Australia, and continues to be Adjunct at the Schools of Psychology at Griffith University, Australia and Victoria University of Wellington, New Zealand.

## Acknowledgements

The authors would like to thank Muznah Siddiqi, AI Research Associate at UNU-CPR for her support during the first stakeholder meeting held in New York in July 2025 (see Appendix D), Alessandra D'Angelo, Digital Specialist at the Asian Development Bank, Marc Lepage, Digital Technology Innovation Specialist at the Asian Development Bank, Ritul Gaur, Digital Technology Specialist at Asian Development Bank and Omar Mohsine, Open Source Coordinator at the UN Office of Digital and Emergent Technologies for their constructive feedback during the different revisions of this report.

# Abbreviations and Acronyms

**ADB:** Asian Development Bank

**AI:** Artificial Intelligence

**AIDPGs:** AI as Digital Public Goods

**CoP:** Community of Practice

**DPGs:** Digital Public Goods

**DPGA:** Digital Public Goods Alliance

**DPI:** Digital Public Infrastructure

**DTN:** Digital Technology Network

**ESG:** Environmental, Social and Governance

**FOSS:** Free and Open-Source Software

**FSF:** Free Software Foundation

**GDC:** Global Digital Compact

**GPAI:** General-Purpose AI Code of Practice

**GPG:** Global Public Goods

**ICT4D:** Information and Communication Technologies for Development

**ITU:** International Telecommunication Union

**LMIC:** Low- and Middle-Income Country

**MDB:** Multilateral Development Bank

**MOF:** Model Openness Framework

**MOSIP:** Modular Open Source Identity Platform

**OSAID:** Open Source AI Definition

**OSD:** Open Source Definition

**OSI:** Open Source Initiative

**OSS:** Open-Source Software

**OSU:** Open Source United

**PII:** Personally Identifiable Information

**RAIL:** Responsible AI License

**SDGs:** Sustainable Development Goals

**STS:** Socio-Technical System

**UN:** United Nations

**UNDP:** United Nations Development Programme

**UNESCO:** United Nations Educational, Scientific and Cultural Organization

**UNICEF:** United Nations Children's Fund

**UN ODET:** United Nations Office for Digital and Emerging Technologies

**UNU:** United Nations University

# Glossary of Key Terms

**AI readiness**
The state of preparedness of an entity (e.g. a country, institution, or organization) to adopt and effectively implement AI technologies. It involves assessing factors such as data quality, infrastructure, workforce skills, governance capacity, and strategic alignment to ensure successful AI integration.[1]

**AI Model**
The combination of architecture (the code specifying layers, connections, input/output types, and learning dynamics) and learned parameters (weights), together with the supporting code needed to instantiate it for inference. Although AI models are essential components of AI systems, they do not constitute AI systems on their own. To transform an AI model into an AI system, additional components, such as a user interface, are required.

**AI Technology Stack**
A collection of technologies, frameworks and infrastructure components facilitating the use of artificial intelligence systems. It provides a structure for developing AI solutions by layering these components to support the end-to-end AI lifecycle.

**AI System**
A machine-based system that, for explicit or implicit objectives, infers from input, how to generate outputs such as predictions, content, recommendations, or decisions that can influence physical or virtual environments. Different AI systems vary in their levels of autonomy and adaptiveness after deployment (OECD, 2024).[2]

**Artificial Intelligence (AI)**
Although numerous definitions of Artificial Intelligence (AI) exist, this report uses, the one adopted by UNESCO in its Recommendation on the Ethics of Artificial Intelligence (2021): “AI systems are systems which have the capacity to process data and information in a way that resembles intelligent behaviour, and typically includes aspects of reasoning, learning, perception, prediction, planning or control.”[3]

**Digital Public Goods (definition provided by DPGA)**
Open-source software, open datasets, open AI systems, and open content collections that meet the DPG Standard, namely solutions must adhere to privacy and other applicable laws and best practices, do no harm by design, and help attain the SDGs.[4]

**Digital Public Infrastructure (DPI)**
Digital systems and platforms that provide secure interactions between individuals, businesses and governments, facilitating services such as digital identity verification and electronic transactions. DPGs can serve as implementations of DPIs.[5]

**DPG Standard Council**
The governing body responsible for guiding the development and maintenance of the DPG Standard. It facilitates the process by reviewing, organizing, and consolidating feedback from the DPG community, ensuring that the standard is developed through a well-planned and transparent approach.[6]

**Environmental, Social and Governance (ESG)**
A framework used to assess how organizations manage environmental impacts, social issues, and governance practices in decision-making, accountability, and long-term performance.[7]

**Free-riding**
Benefiting from a collective good without having incurred the costs of participating.[8, 9, 10]

**Generative Artificial Intelligence**
A category of AI that can create new content, such as text, images, audio, video, or code, in response to prompts or other inputs.[11]

**Global Public Good (GPG)**
Public goods can be considered global when their benefits extend across countries and generations and when it is difficult or impossible to identify a geographically restricted community of beneficiaries. Examples include climate stability, global health security, and knowledge.[12]

**Information and Communication Technologies for Development (ICT4D)**
The use of digital and communication technologies to support social, economic, and institutional development, including improved service delivery and poverty reduction.[13]

**Modular governance**
A mode of regulatory and institutional design in which complex governance challenges are decomposed into discrete, semi-autonomous functional units or modules, each addressing a bounded sub-problem. These units or modules are coordinated through standardized interfaces, shared rules, or platform architectures rather than through a single, monolithic authority. The concept draws on modularity theory from systems engineering and organizational design.[14]

**Non-excludability**
A characteristic of a good whereby it is prohibitively costly or impractical to prevent individuals from using it once the good is provided.[15]

**Non-rivalry**
A characteristic of a good whereby one person's use does not diminish the amount or benefit available to others.[16]

**Open Content**
Content made available under an open license that permits others to access, use, reuse, adapt, and share it, subject at most to conditions such as Attribution or Share-Alike.[17]

**Openness**[18]
A contextual-dependent concept implying access to resources that are otherwise closed or restricted in degrees; it can also refer to a more participatory mode of production.

**Open-Source Software (OSS)**
Software that anyone can freely access, use, modify, and share under a license that complies with the Open Source Definition.[19]

**Open Source AI Definition (OSAID)**
An AI system that is freely available to use, study, modify and share, as defined by the Open Source Initiative. This includes datasets used to train the model, its code, and model parameters, promoting a collaborative and transparent approach to AI development allowing anyone to create a substantially similar result.[20]

**Open Source United (OSU)**
A UN-wide initiative to coordinate and scale open source efforts across the United Nations system by creating shared spaces for UN agencies, funds and programmes to co-develop, share, and reuse open source solutions.[21]

**Public Goods**
Resources available to all and which anyone can utilize multiple times (non-excludable) by anyone without diminishing the benefits they deliver to others (non-rivalrous). The scope of public goods can be local, national, or global.[22]

**Responsible AI**
A broader set of principles that addresses not only the ethical considerations of AI, such as moral values and principles, but also the social, legal, and economic implications of AI. Responsible AI aims to ensure that AI systems operate transparently, comprehensibly, and responsibly. Additionally, it promotes the development and implementation of these systems with a focus on promoting diversity, equity, and inclusion.[23]

**Socio-Technical Systems (STS)**
A design approach to complex organizational development that recognizes the interaction between humans and technology in workplaces. The term also refers to coherent systems of human relations, technical objects, and cybernetic processes that are inherent to large, complex infrastructures.

**United Nations Open Source Principles**
A set of eight guidelines adopted by the UN Chief Executive Board's Digital Technology Network (DTN) that provide a framework for the use, development, and sharing of open source technologies within the UN and globally. The principles are: 1) Open by default; 2) Contribute back; 3) Secure by design; 4) Foster inclusive participation; 5) Design for reusability; 6) Provide documentation; 7) RISE (Recognize, Incentivize, Support, Empower); and 8) Sustain and scale.[24]

# Executive Summary

AI systems are increasingly being positioned as potential Digital Public Goods (DPGs) to accelerate progress towards the Sustainable Development Goals (SDGs). Yet, despite major global commitments, most notably the Global Digital Compact's call to "develop, disseminate and maintain safe and secure open-source software, open data, open artificial intelligence models and open standards that benefit society as a whole", very few AI systems currently meet the DPG Standard in practice. This report explains why, and what must change for "AI as Digital Public Goods" (AIDPGs) to become a credible, implementable pathway rather than an aspirational label.

Commissioned by the Asian Development Bank (ADB) and produced by United Nations University (UNU) in partnership with UN Office of Digital and Emergent Technologies (UN ODET), this assessment combines: (i) a structured desk review of policy, legal, and technical frameworks on DPGs, openness, and AI governance; (ii) key informant interviews with cross sector experts spanning the UN system, governments, civil society, academia, and the private sector; and (iii) a global survey to test whether interview themes hold across a broader sample and to surface where perspectives diverge by region, sector, and AI readiness.

Public institutions and development partners are increasingly seeking AI-enabled solutions that are reusable, interoperable, and locally adaptable. Yet AI systems differ from conventional DPG software in ways that complicate their assessment and stewardship: training data provenance is often opaque or constrained; model behaviour cannot be inferred from the source code alone; and many harms arise at deployment and reuse rather than at release. Without clearer, evidence-based criteria for what must be open (and what can be governed through managed access), and without clearer accountability across the AI value chain, the "AIDPG" label risks becoming unattainable (blocking public-interest innovation) or diluted (undermining trust and safety).

Our evidence points to four conclusions that shape a realistic path forward for AI as DPGs:

• **Openness is multi-dimensional, not binary.** Stakeholders consistently treat openness as a spectrum across components (code, weights, data, documentation) and over time (staged release), while current operational tests skew towards model-release binaries (e.g., OSI's Open Source AI Definition) that do not fully address composite, service-based, or locally assembled systems. While openness cannot be reduced to a single binary test, practical governance requires a common reference framework. Accordingly, this report supports the use of the Model Openness Framework as a shared taxonomy for describing and comparing degrees of openness, rather than as a definitive pass/fail measure of whether an AI system is "open."

• **Societal benefit and SDG alignment are not guaranteed by openness.** "Open" and "public-good" cluster as distinct ideas; privacy and safeguards sit at the centre, linking openness to legitimacy and trust, and making responsible data practices a gating condition for public-sector adoption.

• **Governance must be treated as a lifecycle process, not as a one-time label.** AI systems evolve through fine-tuning, context shifts, and downstream reuse; therefore, "do no harm" cannot be credibly satisfied through licensing alone and requires ongoing testing, documentation, incident response, and mechanisms for contestation and redress.

• **Equity depends on enabling conditions, not on openness alone.** Stakeholders in lower-AI-readiness contexts placed relatively greater emphasis on cost-free access, localization (language and domain fit), and capacity-building; without shared compute, local evaluation capability, and sustainable stewardship models, openness can reproduce rather than reduce dependency and capability gaps.

The report proposes 10 recommendations organized around four priority areas: Standards, Accountability, Finance, and Equity (SAFE). Together, they seek to strengthen the governance of AI Digital Public Goods by promoting greater clarity and transparency, enhancing accountability, supporting sustainable and outcome-oriented financing approaches, and addressing capacity and infrastructure gaps that affect adoption, particularly in developing countries.

# Section 1: Introduction to Digital Public Goods and Artificial Intelligence as Digital Public Goods

## 1.1 Background

The United Nations increasingly recognizes the strategic importance of Digital Public Goods (DPGs), which are open technologies that satisfy the nine indicators of the DPG Standard, including adherence to applicable laws, open licensing, and best practices in support of the Sustainable Development Goals (SDGs), as instruments for global cooperation and digital equity. Critically, the 2019 report[25] of the UN Secretary-General's High-Level Panel on Digital Cooperation catalysed international momentum by calling for a shared digital future grounded in common human values, including human rights, transparency, and inclusiveness.

The recommendations of this panel led to the establishment of the Digital Public Goods Alliance (DPGA) in 2019, a UN-backed multi-stakeholder initiative founded to accelerate the attainment of the SDGs.[26] The DPGA's five-year strategy (2021–2026) aims to equip UN agencies, multilateral development banks (MDBs), and public/private institutions to advance the adoption of DPGs to tackle key development challenges. The DPGA's 2025 DPGs Ecosystem Report[27] underscores the growing number of DPGs and DPGA members. This growth highlights the expansion of available open digital solutions and increasing global participation in efforts to accelerate the realization of the 2030 Agenda and the achievement of its SDGs.

The establishment of the DPGA marked a key milestone in global digital governance. Notably, the DPGA's focus on DPGs for advancing the SDGs[28] provides a critical platform for equitable digital transformation by promoting the development and accessibility of open-source technologies that are free to use, adapt, and scale globally. The contemporary focus on DPGs signals a paradigm shift away from treating digital technologies as market commodities to positioning them as collective assets that support inclusive development and foster international cooperation. The significance of DPGs lies in their capacity to provide scalable, adaptable, and cost-effective digital infrastructures that can be leveraged across contexts, countries, and sectors to promote inclusive and equitable development outcomes.

The Global Digital Compact (GDC)[29], adopted by the UN's 2024 Summit of the Future, affirms the potential of DPGs in delivering services at scale and increasing social and economic opportunities for all. By 2030, it commits to: *"Develop, disseminate and maintain, through multi-stakeholder cooperation, safe and secure open-source software, open data, open artificial intelligence models and open standards that benefit society as a whole."*

The importance of DPGs is also underscored by the prioritization within the UN of operational coordination across agencies in implementing open-source technologies. In late 2024, the UN Chief Executives Board for Coordination established the Open Source United (OSU), a UN-wide initiative designed to coordinate and scale open-source strategy and implementation across the entire United Nations System.[30] OSU aims to break down institutional silos by creating shared spaces for UN agencies, funds, and programmes to co-develop, share, and reuse open-source technologies, aligning on common standards to build solutions that are more efficient, sustainable, and accessible. In parallel, OSU has developed and published the United Nations Open Source Principles, a set of eight guidelines, spanning openness by default, upstream contribution, security by design, inclusive participation, reusability, documentation, empowerment, and sustainability at scale, intended as guidance not only for the UN system but also for endorsement by governments, civil society, and the private sector.[31]

These principles are complemented by a common open-source policy framework and a shared code repository. Taken together, these initiatives signal a maturation of the multilateral open-source ecosystem, moving from high-level normative declarations towards concrete organizational infrastructure for the collaborative stewardship of digital commons. As of 2026, more than 145 organizations have endorsed these principles.

AI systems sit uneasily within all these initiatives. In June 2020, the UN Secretary-General's Roadmap for Digital Cooperation[32] names AI models alongside software, data, standards, and content. Yet the DPG Standard, its nine indicators, and the registry's verification procedures were developed primarily around non-AI software and data artefacts. AI systems differ from these in three respects that matter for DPG assessment: they depend on training data whose provenance and licensing are often opaque or legally constrained; their behaviour cannot be fully specified by source code alone, so openness must extend to neural networks weights, evaluation data, and documentation of training procedures; and the harms they enable are produced largely at the point of deployment rather than at the point of release, which

**Figure 1:** UN Open Source Principles

UN OPEN SOURCE PRINCIPLES

**Open by default**
Making Open Source the standard approach for projects

**Secure by design**
Making security a priority in all software projects

**Design for reusability**
Designing projects to be interoperable across platforms and ecosystems

**RISE**
(recognize, incentivize, support and empower)

**Contribute back**
Encouraging active participation in the Open Source ecosystem

**Foster inclusive participation and community building**
Encourage active participation in the Open Source ecosystem

**Provide documentation**
Providing thorough documentation for end-users, integrators and developers

**Sustain and scale**
Supporting the development

*Source: Linux Professional Institute.*

complicates do-no-harm assessment. Whether and how AI systems can satisfy the DPG indicators considering these differences is a distinct question, and one that has become increasingly urgent as governments and multilateral institutions look to AI to accelerate delivery of the SDGs. In 2025, the DPGA Secretariat updated the DPG Standard for AI systems after convening a Community of Practice (CoP), co-hosted with UNICEF, to develop recommendations for the DPG Standard Council.

The present report addresses this question. Commissioned by the Asian Development Bank (ADB) and produced by the United Nations University International Institute for Software Technology located in Macau in partnership with the UN Office of Digital and Emergent Technology (UN ODET), this study responds to the widening gap between the rapid proliferation of AI systems with potential public value and the need to maintain a rigorous yet practical DPG Standard that is grounded in openness as a core feature of DPGs.

Although the DPG Registry updated the Standard in 2025, there are very few solutions tentatively approved by DPGA, underscoring the distance between normative ambition and operational reality. Drawing on a mixed-methods research design (described in Section 1.6) that combines a systematic desk review, key informant interviews with experts across the UN system, civil society, academia, and the private sector, and a global survey, this report examines what distinguishes AI as DPGs from both conventional DPGs and open source AI more broadly.

## 1.2 Concept of Digital Public Goods

Answering that question first requires clarifying what Digital Public Goods (DPGs) are and where the concept originates from. The DPG Standard, the DPGA registry, and the policy commitments cited above all presuppose a specific definition of public goods adapted to the digital era. Whether AI systems can meet that threshold, and what would need to change in the standard if they cannot, turns on which concept of "public good" is in play.
The next section therefore traces the lineage of DPGs from classical public good theory to the digital commons, before introducing the operational definition and indicators used in the remainder of the report.

### THEORETICAL FOUNDATIONS FROM CLASSICAL PUBLIC GOOD THEORY

The foundational framework of public goods derives from Samuelson's theory of public expenditure published in 1954 (Samuelson, 1954). Public goods are characterized by non-rivalry (i.e. "one man's contribution does not reduce another man's consumption") and non-excludability (i.e. it is prohibitively costly to prevent anyone from consuming the good (Oakland, 1987). Classic examples include national defence, street lighting, clean air, and public parks. (V. Ostrom & Ostrom, 2019) operationalize this into a 2x2 typology of goods, stressing that both characteristics vary in degree rather than being all-or-none.

**Figure 2:** 2x2 Typology of goods

| | Excludable | Non-excludable |
|---|---|---|
| Rivalrous | **Private goods**<br>Food, clothing, cars, smartphones | **Common-pool**<br>Fish stocks, forests, groundwater |
| Non-rivalrous | **Club goods**<br>Streaming services, toll roads, cinemas | **Public goods**<br>National defense, street lighting, clean air, knowledge |

*Source: (E. Ostrom, 1990). Governing the commons: The evolution of institutions for collective action. Cambridge University.*

FROM PHYSICAL TO DIGITAL COMMONS

In the digital context, the concept of public goods has been translated into the concept of Digital Public Goods (DPGs). This concept has a layered history. It has been in use since at least April 2017, when Nicholas Gruen wrote “Building the Public Goods of the Twenty-First Century”.[33] However, this early usage was more of a conceptual framing than a formal definition. The UN Secretary-’General’s High-Level Panel on Digital Cooperation provided the first formal, institutionally authoritative definition of DPGs. The Panel was established by UN Secretary-General António Guterres in July 2018 and published its report “The Age of Digital Interdependence”[34] in June 2019. The Panel argued for greater collaboration around the use of data and development of DPGs to help accelerate achievement of the SDGs by enabling digital technologies to be used at scale and reduce duplication of effort.

In response to the Panel’s recommendation, the Governments of Norway and Sierra Leone, UNICEF and the India-based think tank iSPIRT formally initiated the Digital Public Goods Alliance (DPGA) in late 2019 to operationalize the DPGs definition through a registry and a nine-part standard for vetting DPGs (aka Digital Public Goods Standard). The DPG Standard is an open-source specification (maintained as a collaborative project on GitHub) which sets out the baseline requirements a software artefact must satisfy to receive DPG recognition. The Standard comprises nine indicators, spanning from demonstrated relevance to the UN SDGs to the principle of “do no harm” by design. These indicators address dimensions such as open licensing, the use of open standards, platform independence, data privacy and security, and adherence to applicable laws and best practices.

**Table 1:** The DPG Standard Indicators

| Indicator | Requirement |
|---|---|
| **SDGs Relevance** | DPGs must demonstrate relevance to advancing SDGs. |
| **Use of Approved Open Licences** | DPG must demonstrate the use an approved open licence. For open-source software, only OSI approved licences are accepted. For open content collections, the use of a Creative Commons license is required. |
| **Clear Ownership** | Ownership of assets that the DPG produces must be clearly defined and documented. |
| **Platform Independence** | When the DPG has mandatory dependencies that create more restrictions than the original license, proving independence from the closed component(s) and/or indicating the existence of functional, open alternatives that can be used without significant changes to the core product is required. |
| **Documentation** | DPGs require documentation of the source code, use cases, and/or functional requirements. |
| **Non-PII Data and Content Extraction** | DPGs with non-personally identifiable information (PII) design for possibility of extracting or importing non-PII data and content from the system in a non-proprietary format. |
| **Adherence to Privacy & Applicable Laws** | DPGs must be designed and developed to comply with privacy and other applicable laws. |
| **Adherence to Open Standards & Best Practices** | DPGs must be designed and developed to align with relevant standards, best practices, and/or principles. |
| **Do No Harm by Design** | DPGs must be designed to anticipate, prevent, and do no harm by design. |

*Source: Digital Public Goods Alliance. The DPG Standard Indicators.*

Notably, the Standard is conceived as a living framework: it is periodically refined through community consultation and open feedback processes. The DPGA maintains a registry of verified DPGs and as of the beginning of 2026, the registry contained 241 verified DPGs across four categories:  Open AI System, Open Content, Open Data, and Open Software.

As of late 2025, the Open AI System category boasts only a small number of submissions under tentative review, pending resolution of open questions about training-data openness and do-no-harm evidence. This scarcity reflects the substantial difficulties inherent in assessing AI systems against the DPG framework.

**Figure 3:** Samples of DPGs across the 17 Sustainable Development Goals

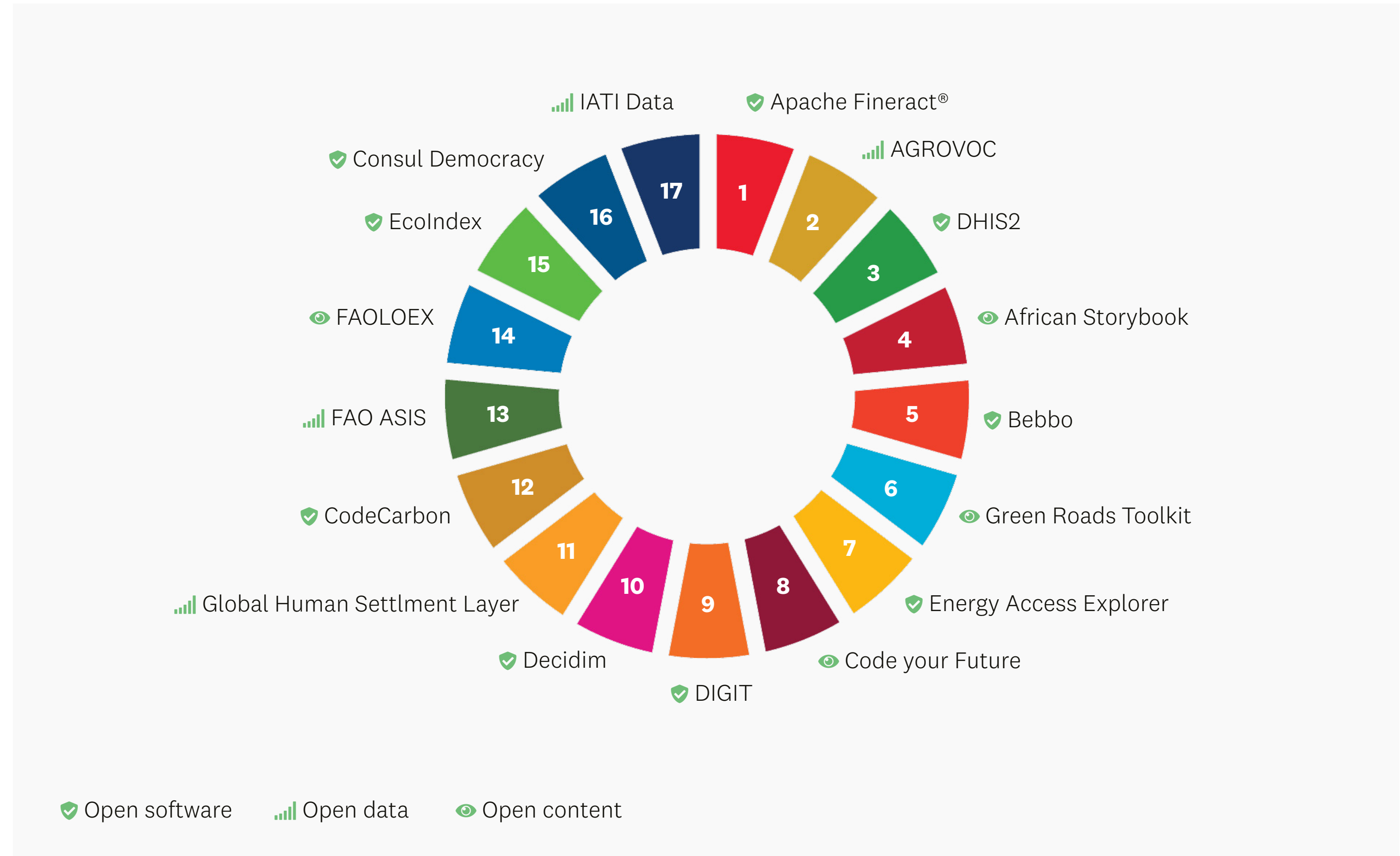

As Figure 3 illustrates, DPGs span the full breadth of the SDG wheel. The distribution visually confirms the empirical observation that open software dominates the certified registry (nine of seventeen selections here), while open data (four) and open content (four) remain comparatively underrepresented despite their equal normative standing under the DPG Standard.

OPEN-SOURCE SOFTWARE

The notion of Digital Public Goods (DPGs) extends a longer tradition of open production, building most directly on the free and open-source software (FOSS) movement that emerged in the 1980s and consolidated through the 1990s. The Free Software Foundation (FSF) grounded its conception of software freedom in four substantive freedoms accorded to users: to run, study, modify, and redistribute source code. The Open Source Initiative (OSI), established in 1998, subsequently codified the Open Source Definition (OSD) along distinct lines, foregrounding developer practice and commercial compatibility rather than the FSF's explicitly political philosophy (Paris et al., 2025). Both currents settled on a shared infrastructure of permissive licensing, source disclosure, and distributed collaboration that would later underpin contemporary articulations of the digital commons (Dulong De Rosnay & Stalder, 2020; Chuang et al., 2022). The translation of this practice into the vocabulary of public goods was largely undertaken in information and communication technologies for development (ICT4D): (Sahay, 2019; Sæbø et al., 2021) argued that FOSS exhibits the non-rivalry and non-excludability characteristic of global public goods, and empirical accounts of platforms such as DHIS2 demonstrated how openly licensed software could function as developmental infrastructure across developing countries (Nicholson, Nielsen, Sahay, et al., 2022). The DPG Standard, stewarded by the DPGA, subsequently operationalized that definition through indicators that preserve 'FOSS's licensing grammar while extending assessment to public-interest considerations such as relevance, privacy, and protection from harm.

## 1.3 Considering Artificial Intelligence as Digital Public Goods

With the growing ubiquity of AI technologies, it has become imperative to guarantee that their applications serve the public interest. The Global Digital Compact (GDC)[35] is the first comprehensive global framework for digital cooperation and AI governance. It aims to promote equitable and inclusive approaches to maximizing the benefits of technologies (including AI) while addressing its associated risks. Moreover, as part of the implementation of the GDC commitments, in 2025, the UN General Assembly established the Independent International Scientific Panel on Artificial Intelligence and the Global Dialogue on Artificial Intelligence Governance. These two mechanisms aim to foster and enhance human oversight of AI systems.

However, while AI systems and their components (models, datasets, weight, code) hold enormous potential societal benefit, their realization as DPGs remains limited. Key technical, ethical, and governance questions remains unresolved. For instance, openness must be carefully balanced with ethical safeguards to address the dual-use dilemma: the risk that AI technologies can be applied for both positive and harmful ends.[36] Moreover, many AI systems often depend on proprietary stacks, commercial data flows, or opaque supply chains (Muldoon et al., 2026) that may challenge “open public good” treatment. The DPG Standard has already begun to address some of these tensions: Indicator 4 (“Platform Independence”) requires submissions to identify and minimize proprietary dependencies, and the Standard Council introduced AI-specific updates in 2025 that fine-tuned Indicators 7 and 9(a) on privacy, responsible data collection, and “do no harm by design”. AI systems exhibit distinctive forms of lock-in which the current operational guidance for Indicator 4 only captures imperfectly. Recognizing this, the DPGA has scheduled a one-year review of the 2025 AI updates for 2026, and the Community of Practice (CoP) on AI systems as DPGs has explicitly framed its recommendations as a basis for further refinement rather than a settled framework. The present report contributes to that ongoing process by examining where AI-specific guidance for the Standard’s existing indicators remains underdeveloped and by suggesting how it might be sharpened ahead of the 2026 review.

Running in parallel to the ’DPGA’s work to extend the DPG definition to AI systems, the Open Source Initiative (OSI) led a complementary effort to define what it means for an AI system itself to be open source. The result, the Open Source AI Definition (OSAID), was released in version 1.0[37] in October 2024 after a two-year multi-stakeholder co-design process (OSI, 2024). Building on the four essential freedoms that underpin the longstanding Open Source Definition for software, OSAID specifies that an AI system qualifies as open source only if it can be used for any purpose without permission, studied and inspected, modified, including to change its outputs, and shared with or without modifications. To make these freedoms exercisable in practice, OSAID further requires access to the system’s “preferred form to make modifications”, comprising the model parameters, the source code used to train and run the system, and sufficiently detailed data information for a skilled person to recreate a substantially equivalent system. Critically, OSAID treats openness as binary rather than as a gradient: a system either grants the four freedoms across these components or it does not. The two efforts are mutually reinforcing: OSAID specifies the openness threshold for AI systems, while the DPG Standard situates that threshold within a broader public-interest frame that also tests for SDG relevance, do-no-harm, and adherence to applicable laws and best practices. Together, they shape the boundary that this report uses between AIDPGs and other forms of partial or restricted release.

**For the purposes of this report, an AI Digital Public Good (AIDPG) is an AI system that meets the DPG Standard’s requirements for open licensing, adherence to applicable laws and best practices, do-no-harm by design, and demonstrated relevance to the SDGs. By AI system, we mean the models, training and evaluation data, code, documentation, and deployment interfaces considered jointly. The term is used here as interim shorthand. As later sections make clear, both its scope (what counts as sufficiently open across the AI stack) and its legitimacy as a distinct category separate from “open source AI” remain contested among the stakeholders this study engaged.**

### 1.4 Findings from the Digital Public Goods Alliance’s Community of Practice on Artificial Intelligence as Digital Public Goods

A helpful starting point for current efforts on AI as Digital Public Goods (AIDPGs) is the work of DPGA and the UNICEF Community of Practice (CoP)[38], which helps structure the key issues for applying the DPG vision and framework to AI. Rather than treating AI as a model release or code alone, the CoP outputs examine the possibility of widening the DPG scope to AI systems and places several linked questions at the centre of the discussion, including how openness should be understood across different layers of an AI system, how responsible AI principles should be reflected in the DPG Standard, how training and testing data should be treated, and how these issues affect the ability of AI systems to serve the public interest and support the SDGs. The CoP discussion paper also approaches the more foundational question of whether AI systems can meaningfully be recognized as DPGs, or whether the more defensible approach is to identify specific AI tools, systems, or use cases that can serve public-good purposes under appropriate governance conditions. This positions AIDPGs not as a settled category, but as an emerging and partly aspirational governance concept linked to wider debates on AI as a global public good. This work is an important foundational step as it moves the discussion away from a narrow software framing and towards a broader socio-technical view in which public value depends not only on code, but also on data provenance, outputs, infrastructure, human oversight, and the wider economic and social context of use.

A discussion paper released by CoP also made clear that the central challenge is not simply whether AI should be open, but how openness should be governed when it interacts with responsibility, privacy, safety, and unequal access to resources. It identified the growing tension between the benefits of openness and the risks that can follow when openness is applied without sufficient attention to data extractivism, data colonialism, community interests, or the unequal ability of different actors to build and benefit from AI systems. The complexity of data within AI systems is a particularly difficult layer, and the paper recognized both the value of open access and the need for more fine-grained governance when datasets are sensitive or carry risks of misuse. At the same time, the paper acknowledged that many of the relevant governance mechanisms extend beyond licensing alone and include documentation, transparency, participatory governance, access arrangements, and resource constraints.

This CoP work initiated the conversation that AIDPG goes beyond technical openness. It also concerns what kinds of openness are appropriate across the AI stack, under what conditions, and with what safeguards. Although the CoP work did not fully resolve these issues, it laid the policy groundwork for further research by identifying three areas that would require more concrete standard-setting. Including the terms under which training and testing data should be opened or managed, the extent to which responsible AI principles should become part of DPG assessment, and the limits of existing licensing models for governing downstream harms.

The subsequent recommendations from CoP focused on agenda items that could be incorporated into the DPG Standard in the near term. The CoP's recommendations significantly advanced the discussion on governance by focusing on specific areas for upcoming assessment, including how openness should be judged, what level of data availability should be expected, how no-harm requirements should be evidenced, and what kinds of documentation could make assessment more consistent across cases. In this sense, the recommendations did not try to address every issue raised in the discussion paper. Rather, they identified the points that were most ready for operational use within the current standard-setting process, while leaving other issues narrowed, deferred, or unresolved, as shown in Table 2. The residual gaps identified in Table 2 should not be read as issues that all fall within DPGA's current mandate or the formal scope of the DPG Standard. Rather, they identify broader AIDPG governance and assessment questions that remain insufficiently addressed across the wider ecosystem.

**Table 2:** Comparison of issues[39] raised in the CoP discussion paper, their treatment in the CoP recommendations, and the remaining gaps

| Topic | CoP Discussion Paper | CoP Recommendations | Residual Gaps |
|---|---|---|---|
| **From AI models to AI systems** | Moves the focus from AI models to AI systems, treating them as STS that include the data layer, output layer, and human-machine interaction layer. | Updating the DPG Standard to assess AI systems, but the concrete proposals are mainly framed around AI model releases and what model developers must provide. | The broader shift to STS is only partly carried through. The recommendations do not translate this system framing into equally clear requirements for deployers, integrators, downstream users, or service-based AI systems. |
| **How openness should be judged across AI components** | Treats openness as a multi-component question across the AI stack and recognizes that openness may differ across components and contexts. It also highlights the ambiguity of what "open" means in AI. | Using a binary test for AI model releases: either conformant or nonconformant with OSI's open source AI definition. It also supports transparent documentation of what is open in each AI component. | The recommendations provide an operational rule for model release, but they do not fully resolve how openness should be assessed for systems assembled from components with different openness conditions. The broader question of differentiated openness across the stack remains only partially addressed. |

| Topic | CoP Discussion Paper | CoP Recommendations | Residual Gaps |
|---|---|---|---|
| **Training and testing data** | Treats data as a central governance issue and connects it to privacy, misuse, collective rights, data extractivism, data colonialism, and unequal access to compute and finance. It signals the need for finer-grained governance of training and testing data. | Making as much data available as possible. Where full openness is not possible, it points to open subsets, samples, synthetic data, or instructions for gated access, together with enough information to support replication. It also notes that views on open training data are still evolving. | The recommendations address practical disclosure options, but they do not turn the broader governance concerns around data rights, extractivism, or alternative stewardship models into concrete assessment criteria. The political economy of data openness remains largely outside the operational recommendations. |
| **Participatory governance and user feedback** | It includes participatory and inclusive design as part of responsible AI and states that systems should incorporate continuous user feedback loops. It also references participatory governance as one of the broader mechanisms for opening AI systems. | Not covered in the recommendations as a distinct requirement. | One of the clearest gaps. The discussion paper treats participation and feedback as part of responsible AI, but the recommendations do not translate them into any specific standard, evidence requirement, or governance mechanism. |
| **Accountability, redress, and verification** | Calls for clear accountability and responsibility, clear lines of responsibility for outcomes, and accessible mechanisms for addressing errors or unintended consequences. It also states that public verification or conformance would be preferable to self-declaration. | Providing evidence of testing and mitigation, an AI risk assessment, a responsible use guide, and safety-by-design planning. | The recommendations strengthen what applicants must submit, but they do not provide a pathway for establishing an independent verification mechanism, a conformance process, or a clearer redress architecture. The credibility problem of self-reporting, therefore, remains unresolved. |
| **Downstream use and post-release harms** | Distinguishes between implementation-based harms and use-based harms and raises the problem that open-source developers often lack visibility into downstream use and may not see themselves as liable for later harms. | Excluding responsible AI licences such as RAIL in the DPG Standard, on the grounds that they are not sufficiently enforceable and are not fit for preventing harm. | The recommendations answer one narrow question, i.e., whether RAIL-style licences should be accepted, but they do not provide an alternative framework for governing downstream use, monitoring post-release harms, or allocating responsibility after reuse and adaptation. |
| **Responsible AI and do-no-harm requirements** | Frames responsible AI broadly around fairness, privacy and security, accountability, transparency, and human oversight. | Adding evidence of testing for bias, fairness, security, resilience, transparency, and accountability, together with mitigation measures where harm is identified. It also recommends measures to explain outputs, ensure human oversight, require an AI risk assessment, a responsible use guide, and a safety-by-design plan, with templates to support implementation. | These are still mainly developer-facing due diligence requirements that do not fully address how responsibility travels across the wider AI value chain after deployment. |

| Topic | CoP Discussion Paper | CoP Recommendations | Residual Gaps |
|---|---|---|---|
| **Resource inequality and smaller teams** | Links AI openness to unequal access to compute, finance, and infrastructure, highlighting the need for more equitable AI infrastructure in disadvantaged and low-resource settings. | Acknowledge that some proposed requirements, especially environmental and mandatory documentation requirements, may burden smaller teams and under-resourced contexts. | The recommendations do not specify how the AIDPG Standard should be designed under conditions of unequal compliance capacity. They recognize the burden problem, but do not propose proportionate requirements, differentiated reporting tiers, or support mechanisms to prevent standards from favouring already well-resourced actors. |
| **Environmental impacts** | Treats sustainability as part of the responsible AI discussion. | Adding a carbon footprint estimate for model training as a mandatory requirement and notes that energy consumption should also be considered. It also acknowledges that such metrics can disadvantage smaller developers and depend heavily on local energy mixes and hardware. | It does not yet constitute a comprehensive lifecycle or deployment-level environmental assessment framework for AI systems, nor does it provide any tools or standards for such assessment. |

On openness, the recommendations favour a binary assessment for AI model releases based on conformity with the OSI's definition, while still recognizing the value of transparent documentation on the openness of individual components. Regarding data, the recommendations stop short of demanding full release of all training data and instead suggest providing as much data as possible through open subsets, samples, synthetic data, or instructions for gated access, along with enough information to enable replication. On responsible AI, the recommendations do not support responsible AI licences as the main governance method, particularly because RAIL-style licences rely on use restrictions that are not conformant with the OSI definition of open source. Instead, the recommendations focus more on stronger do-no-harm requirements, such as evidence of testing for bias, fairness, security, resilience, transparency, accountability, explainability, human oversight, and use case risk assessment. The CoP recommendations also suggest using templates to ease implementation and improve comparability across submissions.

Importantly, the CoP recommendations make clear that some issues remain open rather than settled. Most notably, they recognize that views on open AI training data are still evolving and call for further work with stakeholders to develop norms, standards, and governance models for responsible data sharing in the public interest. Moreover, the CoP recommendations did not seek to settle all the broader questions raised in the CoP discussion paper, particularly those around responsible data sharing, public verification, participatory governance, downstream use, and the practical implications of the potential AIDPG Standard for smaller teams and resource-constrained contexts. This leaves important space for further work on how AIDPG can be assessed in ways that are both practically usable and normatively robust, especially where system-level governance questions remain less developed.

Building on the CoP discussion paper and the subsequent recommendations, this report focuses on the areas that were narrowed, deferred, or left unresolved across the CoP process, particularly the shift from models to systems, component-level openness, participatory governance, public verification, downstream use, and the compliance burdens faced by lower-resource actors, while also extending the analysis to broader gaps that neither document addressed in full.

## 1.5 Research Gaps Addressed in this Report

A systemic expansion of AI systems as DPGs requires clearer standards, ethical guidance, and a more specific assignment of responsibilities to help it develop from a promising idea into a credible supporting tool for sustainable development. Existing work has shown that openness in AI is not a single condition but a set of choices across data, models, code, documentation, access, and governance (OSI, 2024; UNICEF, 2023), yet it still falls short of offering practical criteria for bridging DPG principles as defined by the DPG Standard with the technical and legal realities of AI systems. As of 2025, the DPGA Secretariat has not yet identified a single AI system that fully meets the DPG Standard, and only a handful of submissions are close to compliance (DPGA, 2025).

This near-absence of verified AI systems as DPGs, despite considerable international momentum and the inclusion of AI in the original UN Roadmap for Digital Cooperation, signals that the current requirements, particularly the mandate for fully open training data and open weight, may be misaligned with the practical constraints under which most public-interest AI systems are developed. It is a critical gap because many AI systems designed for social benefit cannot fit into a strict all-or-nothing model of openness, especially when full release of training data conflicts with privacy, community rights, safety, or legal obligations (Solaiman, 2023).

Forcing diverse AI systems into a binary notion of openness may limit rather than enhance societal benefit. The main research gap, therefore, is the absence of evidence-based standards for deciding what must be open, what can be shared through managed or gated arrangements, and what forms of documentation, stewardship, and justification are sufficient across different layers of the AI stack.

A related gap concerns ongoing governance after release. Once licensing is recognized as too weak to prevent harmful downstream use, more work is needed on how obligations should be distributed across developers, data stewards, funders, deployers, and users, and on what forms of independent verification can make AIDPG claims credible beyond self-reporting. Addressing this question demands multi-stakeholder engagement that goes beyond expert-driven and developed countries-centric deliberation. Such deliberation tends to be conducted primarily in English and convened by foundations and standard-setting bodies from the developed countries, and it presupposes technical literacies, including access to compute, fluency with model documentation, and licensing expertise, that are themselves unevenly distributed across regions. Public interest theory emphasizes that AI systems claiming to serve the common good require public justification, equality of consideration, and genuine deliberative processes involving the communities they affect (Züger & Asghari, 2023). Without inclusive governance mechanisms, there is a risk of "participation-washing" (Sloane et al., 2022) in which formal openness masks persistent exclusion of affected populations from important decision-making.

A further underexplored gap concerns local contextualization and the specific needs of the developing countries. Current AI training datasets carry well-documented structural biases (either linguistic, cultural or geographic) that reflect the dominance of developed countries data sources and software development and that do not align with fairness and inclusion principles that are explicitly outlined in the SDGs and the DPG Standard. When deployed in developing-country contexts, these systems risk displacing local knowledge and reinforcing existing inequalities rather than addressing them. The experience of scaling DPGs such as DHIS2[40] illustrates the paradox: expanding a platform to serve a global user base (macro-level decontextualization) can work against the needs of users in particular locations who require deep recontextualization, and the priorities of international donors may overshadow the voices of communities in remote settings (Nicholson, Nielsen, Sahay, et al., 2022).

Research is therefore needed on how AIDPG frameworks can support, rather than merely permit, local adaptation, including the creation of culturally and linguistically representative datasets, community-driven fine-tuning practices, and governance arrangements that give developing countries actors genuine co-determination over the systems designed to serve them.

More research is also needed on implementation pathways that do not assume high levels of compute, legal capacity, and reporting capacity, because otherwise AIDPG Standard may end up favouring well-resourced actors over smaller teams working in public-interest settings with limited resources.

Taken together, these gaps shape the questions this report sets out to answer:

1. How should openness be assessed across AI system components? What forms of partial, tiered, or managed openness can still potentially qualify an AI system as a DPG?
2. What governance arrangements, beyond licensing alone, are needed to sustain public-interest stewardship after release? How should responsibility be distributed across developers, data stewards, funders, deployers, and users?
3. How can AIDPG frameworks support rather than merely permit adaptation to developing countries contexts, including local-language data, community co-determination, and culturally situated evaluation?

The mixed-methods design described in Section 1.6 was selected to answer these questions by combining a desk review of existing frameworks and literature, key informant interviews that surface tacit knowledge from practitioners, and a survey that tests whether interview themes hold across a broader cross-sector sample.

The study is distinguished from prior work in four respects:

1. **Conceptual Framing:** Our specific approach to defining and operationalizing the concept of AIDPGs.
2. **Comparative Regional Analysis:** Our deliberate comparison of regional perspectives (e.g. developed countries vs. developing countries) to identify areas of consensus and divergence, thereby informing targeted regional policies.
3. **Mixed-Methods Stakeholder Analysis:** The integration of qualitative and quantitative data to assess stakeholder perspectives, highlighting existing opportunities, strengths, and barriers regarding AIDPGs.
4. **Multi-level Governance Examination:** The comprehensive analysis of governance structures at various levels (e.g. at national-, macro-, and product-levels) within a single study to inform an overarching policy framework.

**The findings of this report will be developed as non-binding recommendations for the wider DPG ecosystem, including funders and donors, governments, and implementing agencies, as well as for the DPG Standard Council on matters within its remit.**

## 1.6 Methodology Overview

The residual gaps identified above – the shift from models to systems, component-level openness, participatory governance, independent verification, downstream use, and compliance burden on lower-resource actors – are not the kind that can be closed through further desk analysis alone. They require both the tacit judgment of practitioners working on AIDPG design and deployment, and a test of whether emergent positions hold across sectors and regions. To address this matter, the study employs a sequential mixed-methods design, beginning with expert consultations and a desk review to establish foundational knowledge, followed by key informant interviews, and targeted surveys.

An initial expert consultation (see Appendix D) was conducted to inform and refine focal factors of the ecosystem of DPGs, which identified governance and the perpetuation of inequality as critical areas of focus. By mapping the opportunities and barriers for AI systems to qualify as DPGs, and elucidating the key actors and drivers in the governance of AI systems and DPGs, this initial consultation shaped the parameters of the subsequent desk review, interview protocols, and survey design.

The desk review serves two critical functions:

1. Mapping the conceptual landscape at the intersection of AI systems and DPGs, DPGs and open source AI through analysis of academic literature, policy reports, licensing frameworks, and case studies.
2. Identifying knowledge gaps to refine research instruments.

Subsequently, semi-structured interviews[41] were conducted with purposively sampled AI knowledge experts to capture informed perspectives on critical issues surrounding AIDPGs. Following the interviews, key themes were identified based on the core factors and dynamics that participants highlighted as most critical for AI systems and DPGs.

To quantify these qualitative trends, a survey was designed utilizing standardized scales to systematically measure participant agreement. The survey was then disseminated to a broader network of AI knowledge experts across different sectors.[42] Collecting these ratings allowed for the assessment of consensus levels and the analysis of correlations across a wide set of variables, serving to validate and extend the initial interview findings on a larger scale.

Both the interview and survey samples of this study is diverse and balanced, purposefully designed to capture cross-sectoral perspectives and address the focal issue of inequality raised during expert consultations. To ensure equitable representation, participants were drawn from both the developed countries and developing countries, representing regions with varying levels of AI capacity and economic development (see Figure 4 for a breakdown of survey participant distribution by sector).

**Figure 4:** Survey Participant Distribution by Sector
(N = 37)

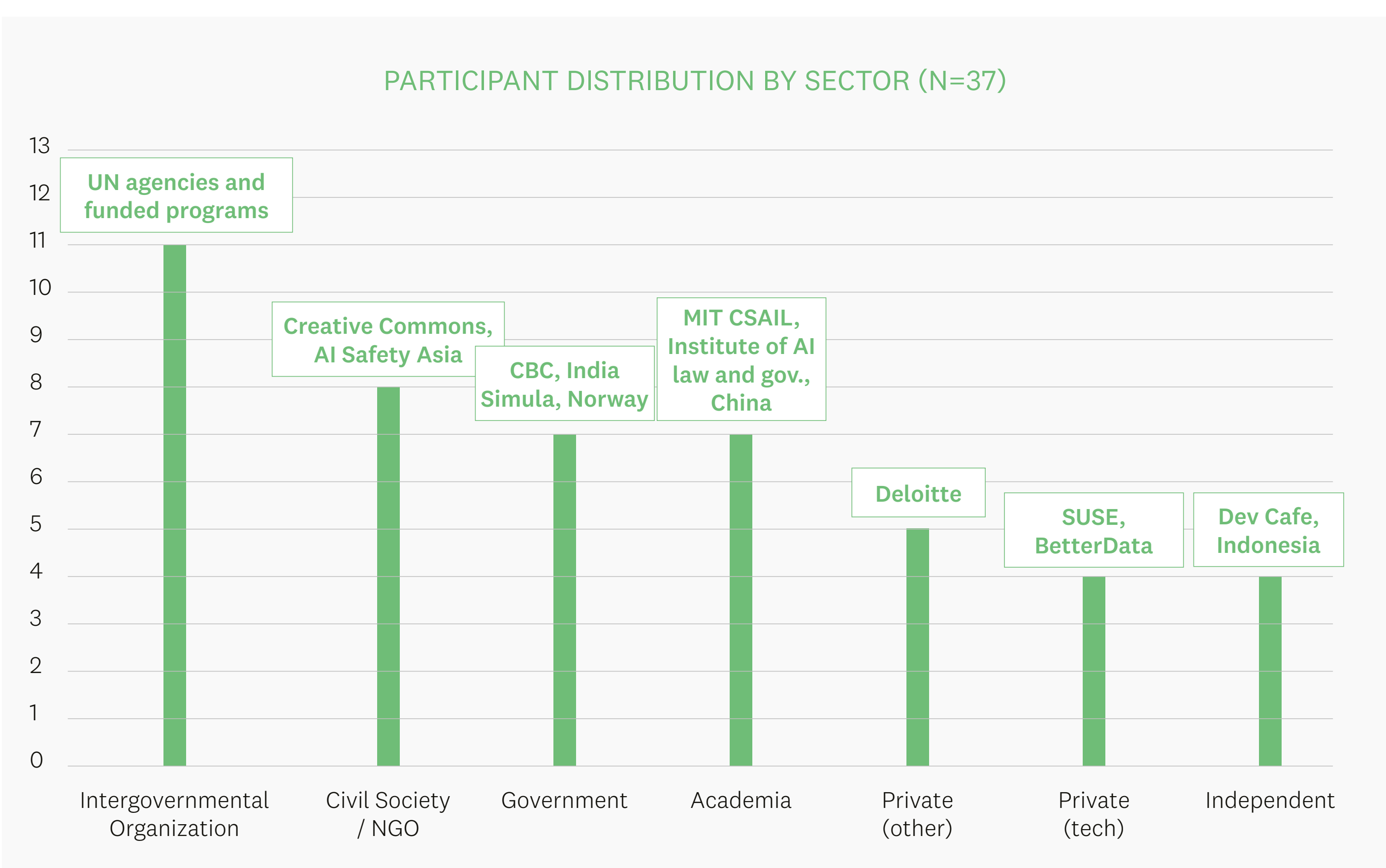

To effectively and reliably compare these diverse perspectives, stakeholders were categorized by their professional sector and geographic context. Geographic locations were mapped against three key frameworks: the developed countries versus developing countries classification[43], country income levels[44], and AI readiness levels.[45]

This balanced, iterative approach provides a robust foundation for drawing meaningful cross-regional and cross-sectoral comparisons, ensuring that the resulting policy recommendations are grounded in comprehensive evidence and stakeholder consensus.

Full methods are outlined in Appendices A, B and C.

# Section 2: Defining the Landscape: Scope, Openness, and Licensing

This section establishes the conceptual foundation for the rest of the report. It draws on stakeholder views, literature, and existing frameworks to examine how AIDPGs are defined, where openness sits on a spectrum rather than as a binary, and how licensing choices across code, data, and model weights shape whether an AI system can meaningfully function as a digital public good (DPG).

## 2.1 Stakeholder Views on the Definition of Artificial Intelligence as Digital Public Goods

The DPG Standard, in its 2025 revision, sets out a normative specification of the attributes an AI system must exhibit to be recognized as a DPG. Across the broader academic and multi-stakeholder discourse, however, a fully consolidated conceptualisation of AIDPGs remains an open question, with parallel frameworks, including the OSI's Open Source AI Definition, the Model Openness Framework (White et al., 2024), and the OECD's AI Openness Primer for Policymakers (OECD, 2025), emphasising different components and degrees of openness.

To deepen our understanding of how these different attributes come together, we asked in our survey, stakeholders to rate the essential characteristics of AIDPGs identified during our desk review and interviews in order to examine how these essential attributes interact. This analysis revealed two distinct paradigms

**Figure 5:** Force-directed correlation network demonstrating the strength of associations among essential attributes of AIDPGs

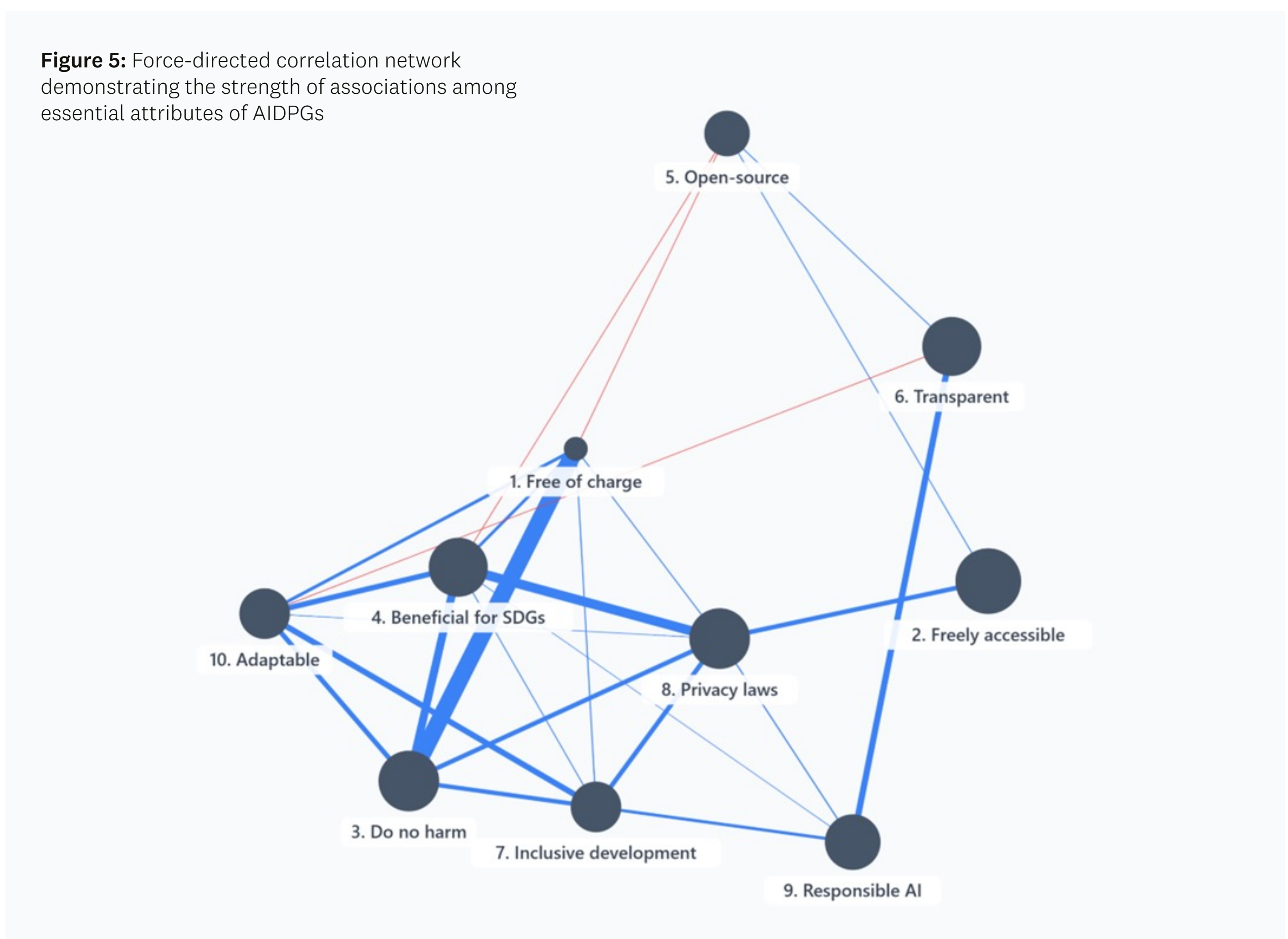

*Note: The node size corresponds to the average rating of each attribute, while the thickness of the connecting red and blue lines denotes the strength of their association. Red lines denote negative correlations, while the blue lines denote positive correlations.*

of AIDPGs that can sometimes conflict, highlighting areas of tension that must be resolved to establish a coherent approach to AIDPGs. Furthermore, to build upon our understanding of these attributes, we asked stakeholders whether a definition for AIDPGs is even necessary, given that a definition for DPGs already exists. This inquiry uncovered significant sectoral differences, revealing additional nuances that must be considered when defining AI as a DPG.

### ESSENTIAL ATTRIBUTES OF AIDPGs: PRIORITIES AND INHERENT TENSIONS

Our survey results are mapped using a force-directed correlation network (see Figure 5). In this visual map, stakeholders rated the importance of each essential attribute of AIDPGs, where positive correlations pull nodes together. Items with highly correlated ratings are connected by thicker lines.

### OVERALL RATINGS AND DEMOGRAPHIC NUANCES

**Figure 6**: Average ratings of AIDPGs core attributes across different regions with dissimilar levels of AI readiness

*Note: Rating ranges from 1 to 5.*
*Abbreviations: SDGs, Sustainable Development Goals; Avg., Average.*

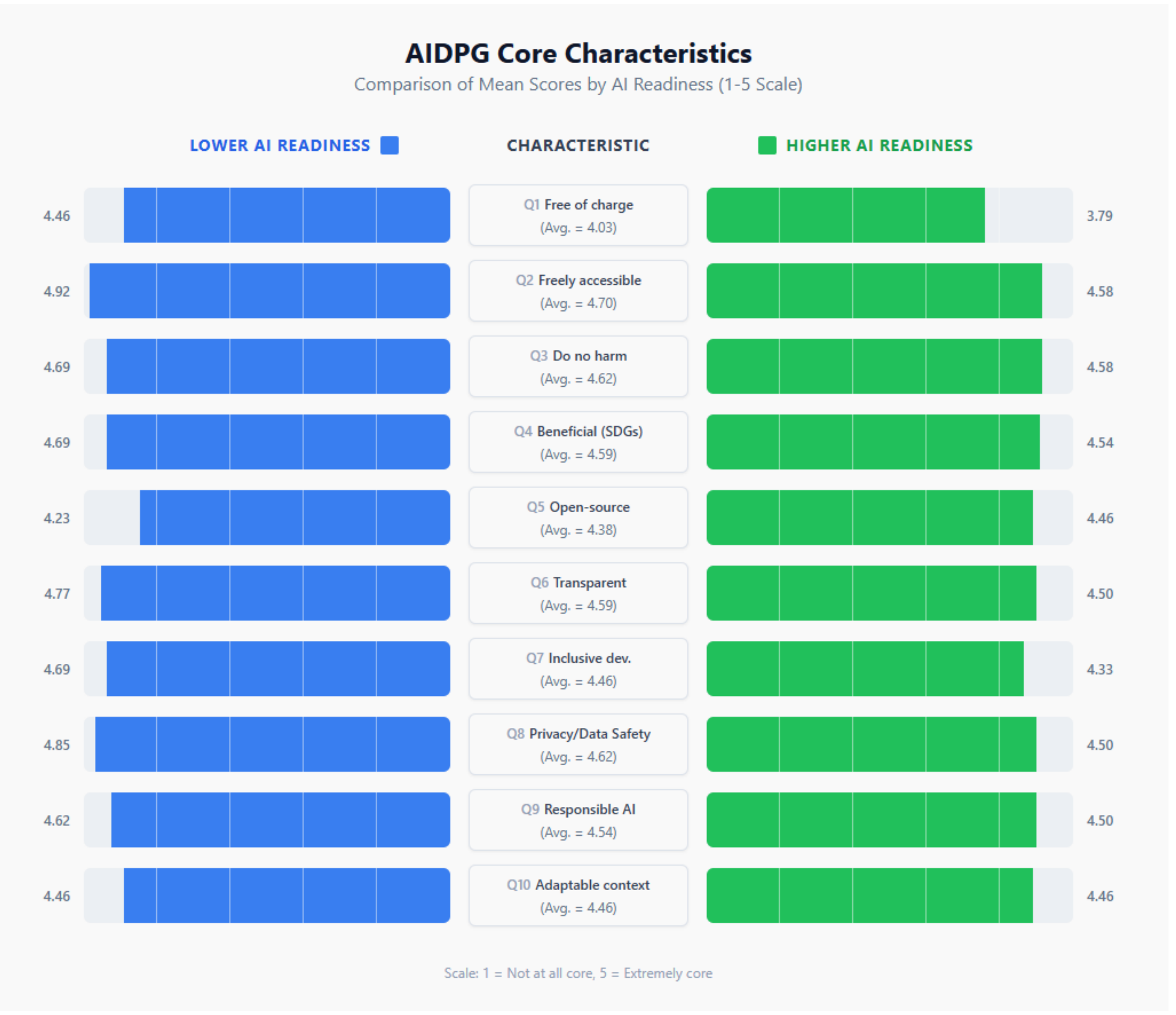

Overall, “freely accessible (i.e. without third-party interference)” emerged as the highest-rated AIDPG attribute (Avg. = 4.70/5), while “free of charge received the lowest relative rating (Avg. = 4.03/5). However, this overall average masks a critical divide: respondents from countries with a lower AI readiness rate being “free of charge” are significantly higher than those from higher AI readiness nations. This highlights that while accessibility is universally valued, eliminating cost remains a much higher priority for less AI-ready regions.

**Attribute Associations and Competing Paradigms**
Beyond average scores, examining the associations between these characteristics using a force-directed correlation network (Figure 5) provides deeper insights into how these attributes cluster together into distinct areas of focus.

The visual map reveals two primary clusters:

(a) A cluster centred around being “beneficial for the public and helping to achieve SDGs”,
(b) A cluster focused on being “freely accessible and transparent”.

This two-cluster structure suggests that while both sets of attributes are recognized as core to AIDPGs, they are perceived as distinct, and sometimes competing concepts. Notably, transparency and free accessibility are weakly positively correlated with public benefit and SDGs, while being strictly “open-source” is negatively correlated. **This highlights a critical tension: the push for maximum openness can sometimes conflict with the mandate to ensure safe, beneficial outcomes for the public. Our findings suggest that this friction must be reconciled if an AIDPG framework is to successfully embody all these characteristics simultaneously.**

**The Central Role of Privacy and Safeguards**
Bridging this definitional divide is the attribute of “complying with privacy laws and safeguarding user data” and “Responsible AI attributes”. In our force-directed correlation network, these nodes are situated at the centre. This indicates that privacy, data protection, and building public trust are recognized as foundational, serving as a critical connecting link that is highly relevant to both the “openness” and “public benefit” clusters.

## DOES AIDPG REQUIRE ITS OWN SPECIFIC DEFINITION?

While high survey ratings and recurring interview themes reveal a general consensus on the core attributes, there are diverse views on whether the unique nature of AI necessitates a strictly separate definition from standard DPGs, or merely an evolution of existing frameworks. When asked explicitly if AIDPGs require their own specific definition, 20 interviewees were divided: 45% (N = 9) argued definitively for a separate definition, 20% (N = 4) opposed it in favour of using existing frameworks, and 35% (N = 7) took a nuanced middle ground, suggesting an evolution of current definitions to include AI-specific criteria. Direct quotes are integrated to illustrate these varying perspectives, accompanied by the interviewee’s identifier (followed by a number to protect their anonymity), professional sector, and geographic region.

**Synthesis of Key Trends in Stakeholder Interviews**
Most interviewees agree that AI presents unique challenges that the existing DPG definition does not fully capture. The largest group of participants (45%) argued definitively for a separate definition, emphasizing that AI is technically distinct from traditional software. It involves non-determinism, opaque “black-box” logic, and complex training pipelines. Furthermore, AI amplifies unique ethical and legal risks (e.g. misinformation and manipulation) that require tailored governance across a complex lifecycle (data labelling, model training, evaluation, and retraining).
**— Interviewee 13** (Technological Development and Civil Society, developed countries)

*“I think there needs to be a more, very specific definition for AI, considering there’s some sensitivities over AI data, training data ... The bias issue with AI model used in Asian countries might not work in Western countries ... There are a lot of complications related to AI.”*
**— Interviewee 17** (Technological Development, developed countries/developing countries)

Conversely, the interviewees arguing against a separate definition (20%) view AI as a continuation of broader digital transformation, akin to the evolution of big data or smart cities. They warn that treating AI as entirely separate risks regulatory fragmentation and overcomplication, suggesting that definitions should remain focused on outcomes and values (e.g. benefit to humanity) rather than the underlying technology.

*“No, I think this whole thing about models is that we’re getting away from the purpose. Most people don’t care what model it is ... **I think what we need is some agreement on what standards are and how the governance will be done** ... Most people in the world, would benefit ... from having something reasonable which does the job and allows everyone to get on with making their lives better ... So the focus on making life better ... is what it should be about.”*
**— Interviewee 3** (Civil Society, developed countries)

A significant portion of interviewees (35%) took a middle ground, suggesting an evolution rather than a revolution of terms. Proponents of this view recommend expanding the general DPG defini-tion to include AI-specific criteria. This approach emphasizes the flexibility needed to accommodate the evolving and intersectional nature of AIDPGs.

*"I think it (AIDPGs) could be encompassed within the general definition of digital public goods as these open-source tools help achieve the SDGs, respect privacy, and use open standards and best practices ... I think everything could still be encompassed within the existing DPG definition. But I think there needs to be some exploration on what's the intersection with public interest, AI, open source, and other types of AI for good that have not necessarily this level of openness."*
**— Interviewee 11** (Technical Development and Policymaking, developed countries)

**Sectoral Differences in Perspectives**
The debate over whether to establish a separate definition is influenced by the professional sector of the stakeholders.

**Table 3:** Sectoral views on the relevance of a separate AIDPG definition

| Sector | Perspectives on AIDPG definition |
|---|---|
| **Technical Developers** | • Favouring a **separate definition**.<br>• Emphasize technical openness, licensing, bias mitigation, and auditability.<br>• As AI systems are fundamentally different from a traditional software, developers argue they require entirely new governance tools, licences, and technical standards. |
| **Policymakers** | • **Mixed views.**<br>• Primary focus is on governance, inclusive development, and human-centric values; split view on the mechanism:<br>o Some see value in unified frameworks to **simplify governance**.<br>o Others call for context-sensitive definitions that can **bridge local and global needs**. |
| **Academia & Civil Society** | • Offer the **most nuanced** perspective.<br>• Focus heavily on social justice, power dynamics, and institutional accountability.<br>• Some **warn against "AI exceptionalism"** and advocate for integrating AI into broader digital governance structures.<br>• Others view the push for a **redefinition as a discursive opportunity** to stimulate debate, uncover ethical dimensions, and drive policy innovation. |

## 2.2 The Spectrum of Openness

Openness is a foundational enabler of scientific progress and technological innovation. In the domain of AI, this logic of cumulative innovation is especially important: current neural network approaches extend concepts developed in the 1950s by pioneers like McCulloch and Pitts. Each architectural breakthrough, from recurrent neural networks to transformers, has depended on the open availability of prior algorithmic, computational, and empirical contributions. In machine learning, the public availability of model components facilitates external evaluation and accountability, accelerates cumulative research, lowers barriers to entry for developers and institutions with fewer resources, and strengthens public trust through **transparency**. Openness also serves as a mechanism for building trust between organizations and their stakeholders, including regulators, civil society, and affected communities by enabling independent scrutiny and public validation of the claims that AI producers make about their systems (Augspurger et al., 2023).

### LITERATURE REVIEW ON OPENNESS IN AI

A central insight in the recent literature is that openness in AI **is not a binary attribute but exists on a spectrum**, from fully closed to fully open. Paris et al. (Paris et al., 2025) identify 98 distinct concepts of openness in different domains and develop a taxonomy of openness in AI comprising three themes: interactivity, freedom, and inclusiveness, arguing that the narrow open-source software (OSS) framing "introduces assumptions of certain advantages" not fully transferable to AI. Chuang et al. (2022) ground this in the broader history of openness as contextual and multi-dimensional, encompassing access, interoperability, participation, and redistribution rights.

To bring conceptual rigour to this space in the context of AI, (White et al., 2024) propose the Model Openness Framework (MOF), a three-tiered classification system: Open Model, Open Tooling, Open Science, that evaluates AI models against both **completeness** (the availability of all lifecycle AI artefacts) and **openness** (the use of OSI-approved or equivalent permissive licences for each component). The MOF identifies 17 components across the model development lifecycle: model architecture, weights, training code, inference code, training data, validation data, test data, benchmark data, research paper, model card, data card, technical report, evaluation results, sample code, datasets for fine-tuning, and usage documentation. This approach is described as **more meaningful than a calculated index**, as it guides the model's producer to move from the lower class to a higher class by providing more components released under open licences.

There is also clearly a **need to measure openness beyond technical criteria**. For example, (Schrepel & Potts, 2025) provide a comprehensive evaluation of AI foundation model licences that extends beyond technical criteria to encompass legal, economic and social dimensions. They argue that AI licenses operate as a "bottleneck" whose openness directly influences the flow of knowledge into the commons. Applying their methodology to GPT-4, Llama 3, Gemini, Mistral 8×7B, and Midjourney V6, they find significant differences from existing technically-focused rankings.

## STAKEHOLDER VIEWS ON THE SPECTRUM OF OPENNESS

The analysis of our interviews with knowledge experts reveals that openness is multidimensional, existing across multiple interrelated spectrums. While stakeholders did not explicitly label their approach as "non-binary", their discussions consistently highlighted varying degrees, timelines, and philosophies of what it truly means for AI systems to be open. This reveals a clear, underlying consensus: openness cannot be treated as a simple binary concept (i.e. either "open" or "closed"). Instead, our findings demonstrate that openness is a multi-faceted characteristic, with three distinct dimensions of the openness spectrum emerging from the data. Direct quotes are integrated to illustrate these varying perspectives on openness.

### Extent of Accessibility: Basic Accessibility vs. Full Transparency

The interviewees highlighted that openness can be seen as a spectrum across the extent of accessibility, starting from mere usability and extending to complete visibility of the AI pipeline.

• **Basic Accessibility (Free to Use):** At the foundational level, openness means the tool is free of charge and is available for people to use without financial barriers.

*"Yes, most of the time (AIDPGs) should be free of charge … and not harmful at all to society or whoever wants to use them. So that's the meaning of Digital Public Good …(and) if AI is applied for the Digital Public Good."*
**— Interviewee 15** (Policymaking and Government, developing countries)

• **Full Transparency (The Whole Pipeline):** True open source AI requires far more than just basic access or releasing parts of the AI system. It requires complete visibility into how the AI was built.

*"(The) preferred form of making modifications to an AI system includes the weights, the parameters, and the results of the training. You need to have all the codes that have been used to produce those weights. And that includes the training code, the code used to create the data set, and then the data, of course, if you can share, it also should be made available. And if you cannot share, it needs to be described sufficiently so that it can be rebuilt."*
**— Interviewee 13** (Technical Development, developed countries)

### Timing of Accessibility: Tiered or Staged Access

Openness does not have to happen all at once. Rather, it can be implemented in stages to balance access with safety.

• **Staged Release:** One interviewee suggested a tiered approach to openness, where an AI system is gradually released in increasing levels of openness. The process can begin with publishing research papers and high-level architecture, followed by releasing APIs or playgrounds for basic inferencing, and finally culminating in the release of model weights and guardrails once safety is better understood.

*"(We can) … look at things from a like tiered or staged access approach, right? And gradually release the AI system, increasing levels of openness. So that comes down to things like research papers and the actual architecture itself at a top level, and then the actual API that you're using, or the playground using for inferencing, and then, beyond that, the model weights, guardrails and so on."*
**— Interviewee 5** (Technical Development, developed countries)

### Openness as a Public Utility vs. No Third-Party Interference

The interviewees revealed a philosophical tension regarding what "open" actually means in practice. This pits the idea of AI as a managed public utility[46] (universally provided but with central safeguards) against the strict open-source ethos of absolute freedom from interference.

• **The Public Utility Perspective (Implies Third-Party Control/ Provision):** Several interviewees argued that open AI systems should be treated like a public utility. In this view, "open" means universally guaranteed access. As a public good, it should benefit everyone, and its access should not be hindered by socio-economic barriers or physical limitations (Interviewee 7). To guarantee this level of universal, barrier-free, and safe access, some argue that AIDPGs need to be hosted and regulated by a central third party, such as the government or a non-profit organization.

(In response to a question related to the governance structure of AIDPGs)
*" … We also need to go back a little bit to what's the definition of a public good, like a railway, the airport, or the water or electricity services, those are the traditional public good. So when I talk about the digital public good, which means the people, everyone*

*can use it, and they can be benefit from this. They don't need to because they are poor or disabled, and they have the limitation to access that kind of service. So, how to build the people's trust back to the question? So people need to know that's the kind of service alongside their daily life. It's provided by government, guaranteed by government, and leave no one behind."*
**— Interviewee 7** (Policymaking, developing countries)

- **The "No Third-Party Interference" Perspective (Strict Open-Source Ethos):** In direct contrast, some interviewees argue that true open-source philosophy dictates that users should have absolute freedom to use and modify the asset without asking for or needing any third-party permission. In this view, adding Terms of Use to an AIDPG could compromise its openness, because it places a burden on the user to consult the rightsholder to verify legal compliance and potential liabilities (Interviewee 13).

*"Let's talk about why I answered that the traditional open-source licences and legal framework on AIDPGs needs to be the same. We need ... (that) because we ... (have) the absolutely need to preserve the underlying principle of open source, that is, that the user of the artifact doesn't have to engage with any third party, including the owners of the of the IP in order to exercise their rights ... (This) principle is a founding principle of open source. It's what allows for the ecosystem to exist. And if we start saying that inside the Terms of Use, we can add exceptions, like 'Don't break the law', a very reasonable, high-level description of, yes, you can use it, but don't break the law. Then you are immediately placing a burden on the user and the developer to engage with the right holder to ask 'Okay, which law do you mean?'"*
**— Interviewee 13** (Technical Development, developed countries)

### STAKEHOLDER VIEWS ON PRACTICAL BARRIERS RELATED TO OPENNESS

The interviewees identified several critical hurdles to making AI fully open and accessible. These barriers occur across the pipeline of AI development from data gathering to downstream deployment and legal regulation. Table 4 summarizes stakeholder views on practical barriers to openness, divided in three stages:

#### Stage 1: Upstream Barriers of Openness – Data Privacy and Provenance

- **Data Privacy vs. Openness:** Requiring all training data to be fully open could be problematic because models are often trained on sensitive, personal, or copyrighted data. Existing frameworks, notably those from the Digital Public Goods Alliance (DPGA), acknowledge this tension by advocating for open standards that respect privacy limitations, recognizing that not all AI systems can completely open their datasets.

- **Opaque Data Provenance and Legal Risks:** Data provenance of AIDPGs that is not fully transparent or adequately tracked can lead to substantial copyright and legal risks, complicating the ability to license them as open digital public goods safely.

#### Stage 2: Deployment Barriers of Openness – Misuse, Malicious use, and Downstream Control

- **Unpredictable Downstream Use:** Unlike traditional software, which provides a fixed, "deterministic capability", open AI systems are highly malleable. As one interviewee noted, AI's capabilities are "not deterministic" because it can be adapted for various unforeseen applications (Interviewee 4). This adaptability stems from the fact that an AI system consists of multiple adjustable componentsthat can be independently modified. Consequently, once an open-source model is released, it is incredibly difficult to predict or control its downstream trajectory, creating a significant regulatory barrier.

- **The Danger of Fully Open Models without Safeguards:** Furthermore, the high level of malleability of open AI systems introduces a critical vulnerability for malicious exploitation. As threat actors can independently manipulate the AI components, they can easily remove built-in guardrails, circumvent safety alignments, and repurpose the model for harmful objectives. Coupled with advanced AI capabilities, this unrestricted access creates the potential for widespread harm, such as executing sophisticated cyberattacks, facilitating the development of weapons, or accelerating the spread of misinformation (Interviewees 1, 2, 15, 17 and 18).

- **Lack of Education on Appropriate Use:** There is currently insufficient education for government officials and the public on how to use open data and AI ethically, which can lead to inadvertent misuse (Interviewee 16). Without proper training or a "common process" for handling open systems, users may unintentionally abuse these technologies, complicating downstream control.

#### Stage 3: Governance Barriers of Openness - Licences and Legal Regulation

- **Incompatibility with Traditional Copyright:** Traditional open-source software (OSS) licensing is highly effective worldwide because it is firmly rooted in international copyright law. However, as one interviewee explained, the nature of certain AI artifacts (e.g. model weights and parameters) does not fit squarely into existing legal frameworks, including copyright or European database rights (Interviewee 13). As these components may fall outside of copyright protection, applying traditional licences to them becomes legally ambiguous. Without the global standardization provided by copyright law, these licences cannot be effectively enforced on a global scale (Interviewee 13).

- **Untested Responsible AI Licences (RAIL):** While some developers attempt to mitigate malicious use by adopting RAIL that mandate ethical use, these frameworks currently lack legal certainty. As one interviewee highlighted, unlike traditional open-source licences, these newer frameworks have not yet been rigorously tested in courts and can create a "false sense of security" (Interviewee 11). Without established legal precedents or reliable enforcement mechanisms, developers may mistakenly assume that downstream users are actively abiding by the ethical restrictions, leaving the models vulnerable to misuse.

**Table 4:** Summary of Stakeholder Views on Practical Barriers to Openness

| Pipeline | Specific Barrier | Meaning | Practical Consequence |
|---|---|---|---|
| **Stage 1: Upstream** (Data & Provenance) | Data Privacy vs. Openness | Tension between open data mandates and protecting sensitive/copyrighted data. | Prevents fully opening all datasets; requires privacy-respecting standards. |
| | Opaque Data Provenance | Lack of transparency and tracking in training data origins. | Introduces substantial copyright and legal risks, complicating licensing. |
| **Stage 2: Deployment** (Downstream Use) | Unpredictable Downstream Use | AI is highly malleable and non-deterministic; components can be modified independently. | Challenges the prediction or control of the model's trajectory post-release. |
| | Malicious Exploitation | Threat actors can remove guardrails and circumvent safety alignments. | Creates potential for widespread harm (e.g. cyberattacks, misinformation). |
| **Stage 3: Governance** (Legal Regulation) | Lack of Education | Insufficient training for officials and the public on ethical AI use. | Leads to inadvertent misuse and complicates downstream control. |
| | Incompatibility with Copyright | AI artifacts (e.g. weights) do not fit existing international copyright/database laws. | Renders traditional open-source licences legally ambiguous globally. |
| | Untested RAIL Frameworks | RAIL mandate ethical use but lack legal precedents. | Creates a "false sense of security" regarding downstream compliance. |

## 2.3 Licensing Frameworks

Licences define the legal terms under which software, data, and models can be accessed, modified, and redistributed; they, therefore, play a decisive role in determining whether an AI asset can meaningfully be treated as a public good. In the AIDPG context, licensing is not a single question but a layered one, because each AI component carries different legal logics, stakeholder expectations, and trade-offs between openness and safeguards.

Licensing practices differ across AI assets. For software, the DPG Standard requires OSI-approved licences, either permissive or copyleft. Permissive licences (e.g. MIT, Apache 2.0, BSD) allow users to freely use, modify and distribute software, including integration into proprietary products (Ballhausen, 2019). Copyleft licences (e.g. GNU GPL) require that derivative works be distributed under the same licence and that source code be made available (Ballhausen, 2019). For datasets and content, Creative Commons or Open Data Commons licences are typical; however, copyright, privacy, and producer rights complicate "open" claims for training corpora. For models, licensing is the most contested component: many widely adopted releases are source-available or open-weight rather than OSI-open, and there is no settled consensus on how traditional open-source principles apply to model weights as a distinct artefact (OSI, 2024; OECD, 2025).

The OSI Open Source AI Definition (OSAID) 1.0 seeks to translate open-source principles into the AI context by specifying conditions for the study, use, modification, and sharing of AI systems (OSI, 2024). While OSI licences have a well-established role for AI software components, their application to model weights remains a matter of active debate. Weights are not source code: they are learned parameters produced from training data and compute, and the community has not converged on whether, or under what conditions, standard open-source licence mechanics can govern them. Several questions remain open — including whether release of weights alone qualifies a model as "open source", what disclosures about training data and processes are required, and how rights to modify and redistribute translate to systems that are expensive or infeasible to retrain. As a result, leading model releases continue to rely on bespoke, non-OSI "open-weight" or "community" licences, and the boundary between "open source AI," "open-weight AI," and "source-available AI" remains unsettled (Liesenfeld & Dingemanse, 2024; White et al., 2024; OECD, 2025).

The RAIL family of licences (e.g. OpenRAIL, BLOOM RAIL) represents an attempt to reconcile permissive access with ethical safeguards by embedding behavioural clauses that restrict harmful uses. These licences have enabled the release of high-profile open-weight models such as BLOOM, while introducing constraints aligned with public-interest goals—such as prohibitions on surveillance, disinformation, or human rights abuses (BigScience RAIL License v1.0, 2022; Contractor et al., 2022). Although RAIL-style licences are not OSI-approved and accordingly do not satisfy Indicator 2 of the DPG Standard under OSAID-aligned criteria[47], they remain analytically relevant to the AIDPG debate as an early attempt to embed ethical commitments at the licensing layer. Their core limitation, the lack of downstream enforceability (Seger et al., 2023), helps explain why the CoP relocated the "do no harm" function to Indicators 7 and 9 rather than to the license itself. Crucially, their relevance lies in how they operationalize ethical commitments embedded in the concept of AIDPGs. By explicitly prohibiting socially harmful applications and encouraging responsible downstream use, these licences advance key public-interest requirements.

By contrast, widely used community licences (e.g. Llama 3.1 Community License[48]) allow broad use but retain marketing and naming conditions and, for very large platforms, require separate permission. For public-good claims, such terms affect re-distributability, derivation, and integration into DPI/DPG stacks, especially for governments seeking legal clarity.

RAIL licenses can help express ethical commitments and responsible-use expectations, but they are insufficient on their own to govern downstream harms and should therefore complement, rather than replace, broader accountability and oversight mechanisms. Beyond licences, documentation and transparency obligations increasingly define "openness". The EU AI Act mandates technical documentation and training-data summaries for general-purpose AI, with an endorsed General-Purpose AI Code of Practice (GPAI) offering templates for model documentation, copyright diligence, and, for systemic-risk models, safety/security practices. This establishes process openness (ex ante and ex post accountability) as a complement to license openness (Regulation (EU) 2024/1689 of the European Parliament and of the Council of 13 June 2024 Laying down Harmonised Rules on Artificial Intelligence and Amending Regulations (EC) No 300/2008, (EU) No 167/2013, (EU) No 168/2013, (EU) 2018/858, (EU) 2018/1139 and (EU) 2019/2144 and Directives 2014/90/EU, (EU) 2016/797 and (EU) 2020/1828 (Artificial Intelligence Act) (Text with EEA Relevance), 2024; The General-Purpose AI Code of Practice | Shaping Europe's Digital Future, 2025).

Crucially, empirical audits show widespread license ambiguity and missing attributions across training datasets, a risk for any public-good deployment that aspires to lawful, ethical reuse. Complementary studies propose deprecation and sunsetting methods for unsafe/obsolete models, suggesting stewardship obligations that go beyond initial release (Liesenfeld & Dingemanse, 2024; Choksi et al., 2025).

Scholarly debates in leading venues (Choksi et al., 2025; Paris et al., 2025) caution that "open models" often have short half-lives or partial openness – and argue for more granular taxonomies (code-open, data-open, weight-open, documentation-open) and governance tying openness choices to risk and capability. For public-good contexts, these analyses support a fit-for-purpose openness approach aligned with use cases and institutional capacity (Liesenfeld & Dingemanse, 2024; Choksi et al., 2025; Paris et al., 2025). In practice, this means combining OSI-compliant code licenses, clearly licensed datasets with provenance, model licenses that permit modification and redistribution of weights where feasible, and full documentation aligned with the EU AI Act and OECD openness guidance (OSI, 2024; OECD, 2025; The General-Purpose AI Code of Practice | Shaping Europe's Digital Future, 2025).

Table 5 summarizes the principal licence families relevant to AIDPGs, their typical object (code, data, weights, or system), their core permissions and obligations, their OSI status, and the stakeholder preferences identified in the literature and in the stakeholder consultations conducted for this report.

**Table 5:** Summary of license families relevant to AIDPGs

| Licence family | Typical object | Core components (permissions / obligations) | OSI-approved? | Typical stakeholder preferences |
|---|---|---|---|---|
| Permissive open-source (MIT, Apache 2.0, BSD) | Software / code | Broad reuse, modification, and redistribution, including in proprietary products; minimal obligations (notice, attribution). | Yes | Preferred by private-sector and technical-development stakeholders seeking maximum reuse and low friction for integration; endorsed by the DPG Standard. |
| Copyleft (GNU GPL, AGPL) | Software / code | Reuse and modification with a "ShareAlike" obligation: derivative works must be released under the same license, with source code made available. | Yes | Preferred by civil-society and commons-oriented actors who want to preserve downstream openness; sometimes avoided by private-sector integrators due to copyleft obligations. |
| Open data licenses (Creative Commons, Open Data Commons) | Datasets, content | Open reuse with options for Attribution (BY), ShareAlike (SA), and Non-Commercial (NC) terms; do not resolve privacy, consent, or producer-rights questions. | N/A (not software licences) | Supported by data-governance and research stakeholders; developing countries actors emphasize the need for stronger provenance, consent, and community-rights safeguards. |
| Open-weight / community licences (Llama Community, Gemma, etc.) | Model weights | Permit broad use and fine-tuning, but impose naming, branding, acceptable-use, or scale-based restrictions; often require separate permission for very large deployments. | No | Favoured by large model developers seeking reach with retained control; viewed with caution by governments and DPG implementers who require legal clarity and unrestricted redistribution. |
| Responsible AI Licences (RAIL) (OpenRAIL, BLOOM RAIL) | Model weights (and sometimes code) | Permit broad access with behavioural clauses restricting harmful uses (e.g. surveillance, disinformation, human-rights abuses). | No | Supported by public-interest, civil-society, and several developing countries stakeholders as a way to operationalize ethical safeguards; contested by strict open-source advocates and raise enforceability concerns. |
| Equitable / benefit-sharing data licences (NOODL, Esethu) | Datasets, especially community-generated or low-resource-language data | Permit access and reuse while adding context-sensitive obligations such as benefit-sharing, community safeguards, or differentiated conditions for developed countries/developing countries users. | N/A; not software licences | Supported by developing countries, data-governance, and community-rights stakeholders seeking to prevent extractive reuse; may be viewed cautiously by actors requiring unrestricted open data terms. |

Taken together, the licensing landscape makes clear that no single licence check can capture "openness" in AIDPGs. AIDPG assessment should therefore consider licensing in layers, including code, data, weights, and documentation, recognize that OSI approval applies most cleanly to the software layer while its application to weights remains unresolved, and evaluate non-OSI options (notably RAIL and community licences) against the public-interest and SDG-alignment criteria that distinguish AIDPGs from open source AI more broadly.

## 2.4 Conclusion

This section established the conceptual foundation of the AIDPG landscape by examining three interconnected pillars: scope, openness, and licensing. Together, these elements shape how AI systems are understood, debated, and operationalized as digital public goods today. By mapping the scope of stakeholder perspectives, we uncovered the inherent tensions between the core value of open access and the necessity of public safeguards, as well as the ongoing debate over whether AI requires its own distinct DPG definition. Exploring openness as an evolving, multidimensional spectrum revealed the barriers to making AI truly accessible safely.

Finally, the analysis of licensing frameworks demonstrated how these conceptual debates manifest in practice, highlighting that the operationalization of AIDPGs deeply depends on layered legal strategies tailored to distinct AI components, including code, data, weights, and documentation. The section highlights that navigating this complex landscape may require reconciling the strict ethos of open-source accessibility with the necessity of ethical safeguards, ensuring that AI can meaningfully function as a safe, equitable, and beneficial public good.

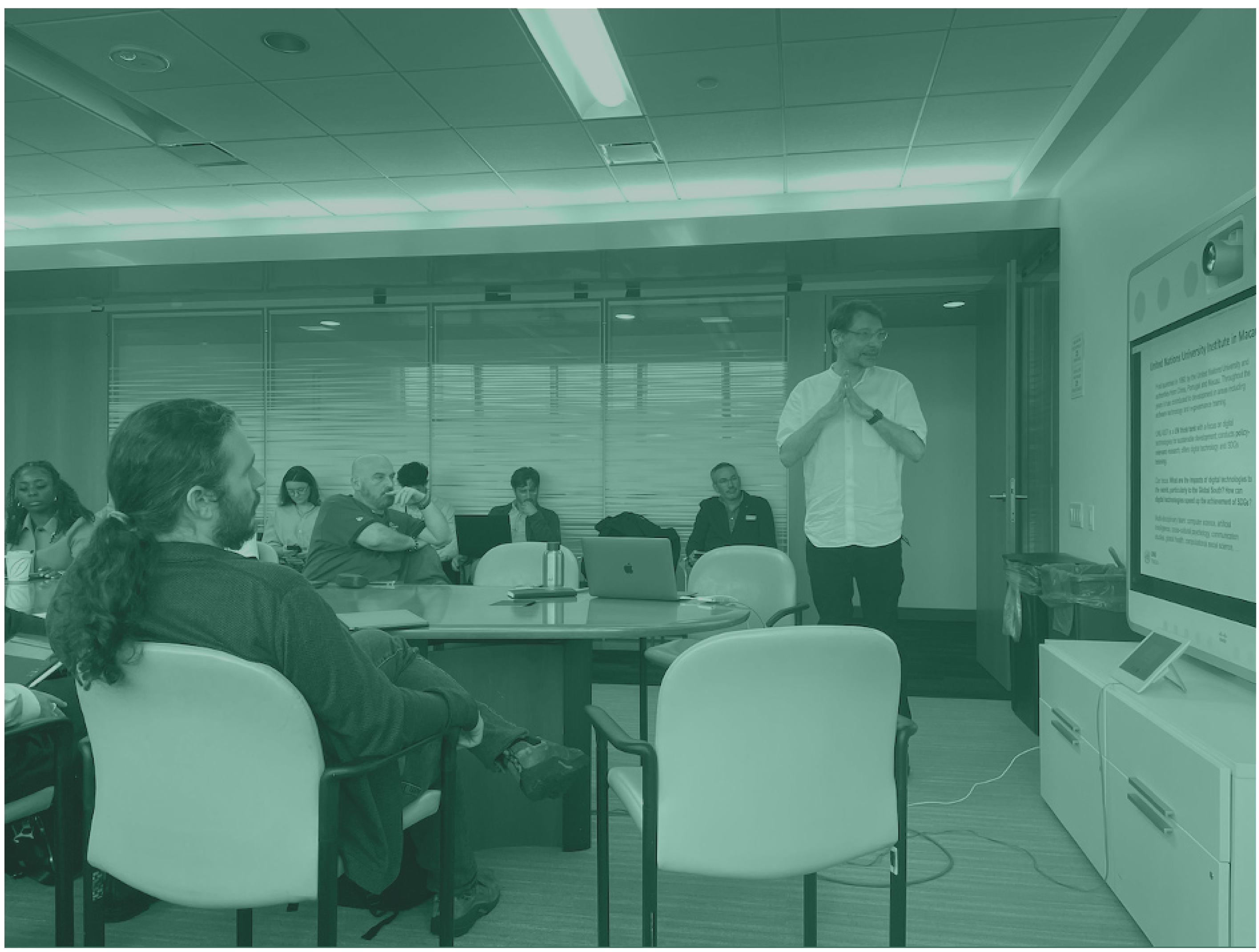

Pictures from Expert Consultation Workshop on Artificial Intelligence as Public Goods

# Section 3: The Benefits and the Risks of Artificial Intelligence as Digital Publics Goods

The case for treating AI systems as Digital Public Goods (AIDPGs) cuts both ways. The same openness that promises equitable access, faster SDG-aligned innovation, and wider participation in AI development is also what introduces new vulnerabilities: weaker data quality controls, broader surfaces for misuse, more diffuse lines of accountability, and unresolved questions about who sustains these systems once the initial enthusiasm wanes. Section 3.1 draws on the literature review and on the interview and survey data to set out what AIDPGs can deliver: cost savings, broader inclusion, and community-led innovation. Section 3.2 turns to the corresponding risks: asymmetries of data and capacity between developing and developed countries, opacity in model provenance, the concentration of effective control among few corporate actors, and the absence of agreed metrics for tracking whether an AI system continues to serve a public purpose once released.

## 3.1 Unlocking Value

LITERATURE REVIEW

Digital Public Goods (DPGs) offer transformative benefits across sectors. By removing proprietary barriers and prioritizing accessibility, they promote equitable access to innovation, accelerate progress towards the SDGs, and foster inclusive development by enabling low-resource communities to leverage advanced technologies (DPGA, 2025). Moreover, DPGs empower global collaboration, allowing diverse stakeholders to co-create solutions for pressing societal challenges in sectors like banking, finance, and healthcare. Approaching AI as a digital public good—rather than solely a profit-driven innovation—allows embedding human-centric values, ethics, and diversity throughout its lifecycle. This perspective emphasizes proactive safeguards and inclusive design to ensure AI development serves societal needs rather than commercial interests alone (AI as a Public Good: Ensuring Democratic Control of AI in the Information Space, 2024). Another critical benefit of AIDPGs lies in the non-excludable and anti-rivalrous nature of "public goods" as defined in economics, whereby their use by one group does not diminish but instead enhances their value for others, and people cannot be excluded from accessing them (Gaurav, 2023). Moreover, AIDPGs foster transparency, ethical governance, and safety in AI deployment, positioning responsible innovation as a shared global priority.

STAKEHOLDER VIEWS ON THE POTENTIAL BENEFITS OF AIDPGS

Stakeholder interviews strongly corroborate these literature findings, emphasizing that AIDPGs drive inclusion, accessibility, and cost savings (Interviewees 1, 2, 3, 5, 15, 16, 17 and 20) while accelerating SDG-aligned services (Interviewees 6 and 20).

*"I think the best (thing is that) it would save time and cost. You don't have to reinvent the wheel. You can piggyback on what is already built and kind of tailor that to your own context and purposes."*
**— Interviewee 17** (Technological Development, developed countries/developing countries)

*"The key benefit is the delivery of services connected to the SDGs. The public good is not the AI. The public good is the SDGs."*
**— Interviewee 6** (Academia, developed countries)

Beyond these established benefits, interviewees revealed interesting insights, particularly highlighting that true value stems not just from open code, but from building sustainable community ecosystems (Interviewees 5 and 13) and fostering local "ownership" to ensure continuous improvement and trust (Interviewees 1, 2, 5 and 8).

*"It's not just open source, it's the community of people around those projects, and developing a sustainable community of people around it, rather than having one super developer that sort of does this right."*
**— Interviewee 5** (Technical Development, developed countries)

*"I would say that ownership would be the main benefit. So people would feel that they own the system in a way. It's part of their community, part of their culture."*
**— Interviewee 8** (Technical Development/Research, developed countries)

Perceived AIDPGs' benefits diverge across regions: developed countries voices primarily value AIDPGs as a vital counterweight to the privatization of knowledge by frontier tech labs (Interviewees 13 and 14), stressing open ecosystems for scientific innovation, interoperability, and regulatory alignment (Interviewees 5 and 13).

*"Given also how science is moving out of the research labs and into private labs, the trend is dangerous. We could get to the point where we do have a privatization of the knowledge."*
**— Interviewee 13** (Technological Development and Civil Society, developed countries)

Conversely, developing countries stakeholders highlight the potential for technological leapfrogging (Interviewees 9 and 10), mitigating digital divides (Interviewees 9, 10 and 12), and achieving cost efficiency for governments (Interviewee 16).

*"Benefits are massive, especially for developing countries. They will be able to leapfrog very quickly to where they could be, because now knowledge is kind of free flow ... That will definitely help many nations."*
**— Interviewee 10** (Private Sector, developing countries)

*"In terms of, I think, resources and costing, so that is very much cost-efficient for the government, and also most of them are willing to share this among each other."*
**— Interviewee 16** (Policymaking and Government, developing countries)

Illustrative Examples: Illustrative examples of these benefits provided by interviewees include MapBiomas[49], which open-sources AI tools for regional biodiversity mapping (Interviewee 11), and rapid language onboarding deployments in Lesotho and Indonesia (Interviewee 5).

Developing countries Highlights: For the developing countries, the promise of AIDPGs lies in reaching last-mile users through localized, domain-specific solutions (e.g. AI models trained in Khmer; Interviewee 15) and overcoming hardware constraints through government-enabled capacity building and context-aware implementations (Interviewees 5 and 17).

*"Cambodian people use Khmer language, and if AI do not provide that kind of feature, it (would be) useless. So there should be a specific, domain-specific AI that allows them to use our own language."*
**— Interviewee 15** (Policymaking and Government, developing countries)

*"Open access to this technology should be combined with the local capacity to actually use it ... You also have to invest in building local capacity to actually use that."*
**— Interviewee 17** (Technological Development, developed countries/developing countries)

## 3.2 Navigating Risks

CHALLENGES OF QUALIFYING AI SYSTEM AS DPGs

Efforts to embed DPG principles into AI systems face significant hurdles, particularly in regions with limited governance over data use and privacy. A major barrier is the lack of accessible, high-quality datasets from the developing countries, which restricts the development of inclusive AI models and perpetuates reliance on data originating from wealthier nations—often misaligned with local realities. For instance, the majority of clinical AI datasets, including those for medical imaging, originate from a handful of high-income settings, predominantly the United States and China (Celi et al., 2022), making diagnostic tools trained on them less effective in African, South Asian, or Latin American healthcare contexts where disease prevalence, demographics, and clinical environments differ.

Additionally, proprietary control over valuable datasets in sectors like finance and healthcare significantly impedes the ability of start-ups or non-governmental organizations to contribute meaningfully to open and equitable AI ecosystems. This uneven data ownership reinforces existing power asymmetries and stifles inclusive innovation. As (Gaurav, 2023) argues, openly accessible datasets and AI models are foundational to the definition DPG, enabling transparency, reproducibility, and broad societal benefit. Yet, the persistent scarcity of such resources undermines the principles of openness and equity, making it difficult to classify many AI systems as true public goods. Without deliberate initiatives to promote data sharing, interoperability, and ethical stewardship, the transformative potential of AI to serve collective interests remains constrained, particularly in low-resource settings where public access to data is most urgently needed.

Moreover, the complexity and opacity of contemporary AI systems further alienate public institutions, particularly in developing contexts where trust and transparency are preconditions for legitimate adoption (Gurumurthy et al., 2022). Empirical evidence indicates that existing explainability tools predominantly serve internal technical audiences, like machine-learning engineers debugging their own models, rather than the end-users, regulators, or policymakers whose decisions such tools are nominally intended to inform (Bhatt et al., 2020).

These challenges are further exacerbated by the dominance of large corporations in AI research and development. The closed nature of their practices, together with their reliance on region-specific datasets, can limit the contextual relevance of AI systems across diverse geographies.

According to DPGA (2025), AI systems shall adhere to the principles of Responsible AI to be considered DPGs. Therefore, robust policies and legislations governing the responsible use of AI are essential for enabling public institutions to harness its full potential while safeguarding user rights and delivering inclusive public services (EY & DEVEX, 2024). Such instruments empower governments to align AI deployment with ethical standards, equity goals, and national development priorities. However, where such instruments are lacking, the transformative promise of AI as a digital public good remains constrained. Without clear regulatory guidance, institutions may struggle to ensure transparency, accountability, and interoperability—leading to fragmented implementation, limited public trust, and missed opportunities to scale AI solutions for social impact. Ultimately, the absence of coherent policy infrastructure risks reinforcing digital divides and undermining efforts to embed AI within equitable, rights-based service delivery systems. Without structural reforms that promote openness, contextual sensitivity, and inclusive governance, AI systems will remain distant from the ideals of a digital public good.

Another emerging threat is the risk of AI-mediated OSS sustainability erosion ("vibe coding") in AIDPG maintenance policy. A recent contribution by (Koren et al., 2026) models the equilibrium effects of AI-mediated software development ("vibe coding") on the open-source ecosystem. Their central finding is that when AI agents assemble software by selecting and composing OSS packages without direct user-maintainer engagement, the visibility and engagement channels through which many OSS maintainers earn returns are eroded. Under their model, if monetisation depends entirely on direct user engagement, widespread vibe coding leads to reduced entry, declining variety, and ultimately welfare loss despite higher short-term productivity. This dynamic weighs more heavily on AIDPGs than on mainstream OSS. AIDPG maintainer communities are structurally thinner and DPG ecosystems serving developing countries contexts compound this contributor risk with the capacity, participation, and financial gaps documented by (Sahay, 2019) and with the scaling asymmetries that tend to entrench rather than close (Nicholson, Nielsen, Sæbø, et al., 2022). Where maintenance already depends on scarce volunteer time and uneven institutional support, the marginal cost of any erosion in the engagement-funding loop modelled by (Koren et al., 2026) is correspondingly higher. For AIDPG maintainers, whose duties extend beyond code to training-data curation, model documentation, and evaluation against locally relevant failure modes (White et al., 2024), this overhead falls on a labour pool least equipped to absorb it, with the risk of deepening the capacity asymmetries that DPG governance is meant to redress.

### ASSESSING AND MEASURING PUBLIC VALUE AND RISKS IN SOCIO-TECHNICAL AIDPG SYSTEMS

Assessing the value and risks of AIDPGs requires moving beyond the individual artifact to the socio-technical system (STS)in which models, data, interfaces, deployment settings, institutions, and users interact. While the current discussions on DPGs and AIDPGs identify important socio-economic benefits and risks, current standards remain less clear on the mechanisms for this assessment in practice. This is a significant gap because openness or declared alignment with SDGs on its own does not show whether an AI system is generating positive outcome, who benefits from it, or whether its risks are being managed over time. Work on public-interest AI suggests that systems intended to serve the public interest need to be examined not only through technical openness, but also through public justification, equality, deliberation or co-design, technical safeguards, and openness to validation (Züger & Asghari, 2023). In a similar way, UNESCO's Ethical Impact Assessment, OECD's lifecycle-based accountability work, and National Institute of Standards and Technology's AI Risk Management Framework all point to the need to assess AI in context, with attention to affected stakeholders, positive and negative impacts, mitigation measures, and ongoing governance across the full system lifecycle, rather than relying on broad claims of social benefit alone (OECD, 2023; Tabassi, 2023; UNESCO, 2023).

In the context of AIDPGs, particularly to increase their competitiveness, clearer standards anchored in measurable indicators across four broad areas are needed.

The first area is the core social value, the extent to which AIDPGs, as systems, improve accessibility, reach underserved populations, and respond to local needs and institutional contexts. The second is the assumed economic value, which entails measurable indicators of whether the openness of AI systems, along with other AIDPG norms, supports long-term cost reduction, reuse, interoperability, local adaptation, and reduced dependence on proprietary vendors. In the DPG literature, the first two dimensions are commonly presented as central advantages of openness and reuse, particularly through faster implementation, lower costs, interoperability, local adaptation, and reduced dependence on proprietary vendors (Martin Meier et al., 2023; Urvashi, 2022). However, literature also makes clear that such benefits cannot always be assumed and depend on supportive local conditions, including implementation capacity, sustainable financing, maintenance arrangements, and a broader enabling digital ecosystem (Martin Meier et al., 2023; Sæbø et al., 2021; Urvashi, 2022). The third area of evaluation should address risks, focusing on whether privacy, bias, security, resilience, misuse, and avenues for contestation and redress can be measured using consistent, comparable indicators suited to the technical and governance characteristics of AIDPGs. Lastly, long-term sustainability should be assessed by examining the extent to which AIDPG systems have viable stewardship arrangements, dependable maintenance capacity, and manageable long-term energy and resource demands. For AIDPGs, this requires attention to the structural and governance realities of OSS, which often depend on distributed contributors, diffuse responsibilities, and limited user-side visibility into maintenance and oversight arrangements. Although this model can support openness, reuse, and local adaptation, it may also weaken clarity around long-term support, funding, oversight, and accountability. Measurement frameworks should therefore be designed to explicitly account for these features.

A useful next step, therefore, is to align AIDPG assessment more explicitly with measurable sustainability and Environmental, Social and Governance (ESG) standards-type approaches in order that claims about public benefit, governance quality, labour conditions across the AI supply chain, and lifecycle environmental performance can be evaluated more credibly and compared more systematically. Clearer measures across these areas would allow decision-makers to compare systems more meaningfully, judge fitness for context, and distinguish between AI that is merely open and AI that delivers durable outcomes under credible safeguards. Such an approach would also help connect product-level AIDPG assessment to broader sustainability and governance reporting, while making the trade-offs and constraints that shape real-world implementation visible. Any such framework would also need to address difficult boundary questions, including how far assessment should extend along the value chain, how accountability should be allocated among actors, and how verification can be operationalized without imposing disproportionate reporting burdens. These boundary questions are especially difficult because corporate AI generally fits within established corporate sustainability and ESG reporting frameworks, whereas AIDPG remains a more ambiguous governance and reporting category.

At the system level, this creates a more visible accountability gap for DPG-like AI systems assembled across multiple actors and components. This challenge becomes sharper when AIDPGs are treated as STS rather than as discrete artefacts. As AIDPGs are increasingly discussed as STS assembled from models, data, services, interfaces, and downstream applications, impact assessment is likely to remain one of the harder design questions in AIDPG governance (UNICEF, 2023). When approached at the STS level, openness, deployment, and public-interest effects in AIDPGs are distributed among somewhat independent actors who do not share a single governance structure, or often no concrete governance structure at all. The DPGA–UNICEF discussion paper already notes that open-source developers may have limited visibility into downstream use and may not view themselves as responsible for harms that arise when their artefacts are integrated into broader systems. For AIDPGs, the boundary of any meaningful assessment should therefore extend beyond the individual component to the larger system in which social, environmental, and economic effects are produced. Emerging work at the intersection of AI, ESG, and sustainability highlights two points that are directly relevant to any such AIDPG assessment. First, existing sustainability-reporting frameworks now offer strong entry points for assessing several impacts that matter for AI systems, including environmental burdens, social effects, and governance-related concerns. Literature shows that AI should be assessed through a broader sustainability lens that adequately accounts for positive and negative effects, interlinkages across domains, and impacts at micro-, meso-, and macro-levels (Sætra, 2021). Structured tools are already available to evaluate and disclose the ESG implications of AI capabilities, assets, and activities, with the aim of improving awareness, governance, and stakeholder communication and treating responsible AI, ESG, and SDGs as mutually reinforcing agendas rather than separate domains (Huang & Yao, 2025; Lee et al., 2025; Sætra, 2023). Moreover, the literature and emerging guidance already identify energy use, emissions, AI labour supply chains, supplier social conditions, and privacy-related risks as relevant areas for AI impact assessment (Sætra, 2021, 2023; Lee et al., 2025; OECD, 2026; Gonzalez-Cabello et al., 2025). Several adjacent impacts relevant to AIDPGs already fall within these established disclosure and due diligence domains.

Second, the literature indicates that ESG-type approaches cannot be imported into AIDPGs without modification. This twofold challenge demands attention to both the existing constraints in AI assessment and the gaps between current AI assessment frameworks and the specific requirements of AIDPGs. For instance, the OECD's work on AI's environmental footprint shows that lifecycle measurement remains fragmented and still requires more standardized indicators, especially for energy, water, and emissions (OECD, 2022). Also, AI-related impacts are often context-dependent, distributed across actors and stages of deployment, and can generate ripple effects over time, making them difficult to capture through static or generic disclosure alone (Sætra, 2021, 2023). This is especially relevant for system-level sustainability concerns such as safety, reliability, resilience, explainability, privacy, fairness, accountability, human oversight, and redress, all of which depend on how the system operates in context (National Institute of Standards and Technology, 2023). This caution applies even more in the case of AIDPGs, where linking use context with development and deployment is not as straightforward as AIDPG systems may be assembled from multiple open and upstream components and where responsibilities for design, deployment, and downstream effects are not always located within a single governance structure (UNICEF, 2023).

AI-specific assessment frameworks are still needed to judge system-level public-interest qualities, while existing sustainability frameworks can anchor adjacent environmental, social, governance, and economic disclosures where recognized categories already exist (Lee et al., 2025; Sætra, 2023; Tabassi, 2023; OECD, 2026). In the AIDPG context, however, the value of such a crosswalk depends on defining the socio-technical boundaries of AIDPGs and clarifying how responsibilities are allocated across upstream, downstream, and integrating actors (OECD, 2026; UNICEF, 2023). Without a clear governance layer, better reporting alignment will still leave open the question of who is accountable for impacts generated across the system (OECD, 2026; UNICEF, 2023).

## STAKEHOLDER VIEW ON THE RISKS OF AIDPGS

Interviews with global stakeholders echo these documented challenges, suggesting that general AI risks such as embedded biases, representational gaps (Interviewees 1, 2, 5 and 20), and the dual-use amplification of harms (Interviewees 1, 2, 5 and 17) are also applicable to AIDPGs.

*"Creating and curating inclusive de-biased training datasets is important because you've got no real control over what people are going to do."*
**— Interviewee 5** (Technical Development, developed countries)

*"Bad actors can also access this technology easily and without, like large code that could actually amplify the negative impact of this. So I think it really amplifies both sides, good and bad."*
**— Interviewee 17** (Technological Development, developed country)

Furthermore, the interviews uncovered additional, nuanced risks specific to open access, notably the "tragedy of the commons", where free-riders, diffusion of responsibility, and insufficient funding threaten the long-term stewardship of projects meant for the global public good (Interviewees 4, 9 and 10), leading to model drift and declining reliability over time (Interviewee 8).

*"One risk is abuse of the system and not being able to improve the public good over time. Because this is like a common. So there's a tragedy hiding behind it. If people use it and do not contribute to it."*

**— Interviewee 4** (Policymaking and Technological Development, developed countries)

*"I would say that the main risk is that the system drifts over time ... So that kind of works fine at the beginning, and then, as more data is in the system, then it starts moving somewhere else, not as expected."*
**— Interviewee 8** (Technical Development and Research, developed countries)

Risk perceptions also diverge significantly across regions. Developed countries experts express concern over the concentration of corporate power (Interviewees 2 and 13), the loss of systemic contestability (Interviewee 12), and malicious generative AI applications like deepfakes and scams (Interviewees 13 and 17).

*"It's related to the market concentrations of power. That is something we need to be very careful with ... So much power will be concentrated in so few companies, all in the developed countries ... dominance of a few actors."*
**— Interviewee 13** (Technological Development and Civil Society, developed countries)

Developing countries voices emphasize the exclusion of low-resource cultures (Interviewee 7) and the lack of sustainable maintenance models without sovereign backing (Interviewees 9 and 10).

*"We don't really have enough concern or enough attention on the people who are living in developing countries. For example, there are some rare languages used by maybe just like a few 1,000 people in the Pacific Islands ... We lack that kind of very rare knowledge which has basically no market value."*
**— Interviewee 7** (Policymaking, developing countries)

*"When we talk about global public goods, you don't have the support of a sovereign nation behind it ... When it comes to digital public goods, yes, because they're still attached to some sovereign nation ... If you want to go truly global public good, long-term viability and maintenance. We haven't resolved that equation yet."*
**— Interviewee 9** (Policymaking, developing countries)

Key examples highlighted in the interviews of these compounding risks include the perpetuation of historic biases in recruitment algorithms (Interviewee 20) and the omission of rare or under-resourced languages due to low commercial incentives (Interviewee 7).

For the developing countries, these vulnerabilities are particularly acute, with interviewees highlighting profound concerns over cultural integrity, where imported AI norms from more digitally mature nations may reshape local values (Interviewee 7). Stakeholders warn that an over-reliance on imported AI systems without concurrent investments in local skills may contribute to a path-dependency that further weakens domestic capabilities and leaves developing communities further behind (Interviewees 16 and 17).

*"Sometimes, when you are very much comfortable with a certain solution, you do not want to step up or upgrade … They are comfortable already, they are contented. They are not aware that there are more sophisticated or high-level solutions … and skills. So they would be left behind."*
**— Interviewee 16** (Policymaking and Government, developing countries)

### 3.3 Conclusion

Across literature and stakeholder evidence, the value and risk of AIDPGs represent two sides of the same openness. Cost savings, inclusion, and SDG-aligned reach materialize only where local capacity, sustainable stewardship, and contextually grounded data are also in place; without them, the same openness can entrench developed countries data dominance, opaque provenance, and corporate concentration, and leave maintenance burdens on the actors least able to absorb them. Realizing the public-interest promise of AIDPGs, therefore, depends less on declaring a system open than on building the assessment, governance, and resourcing arrangements that translate openness into durable impact. The next section turns to those governance arrangements.

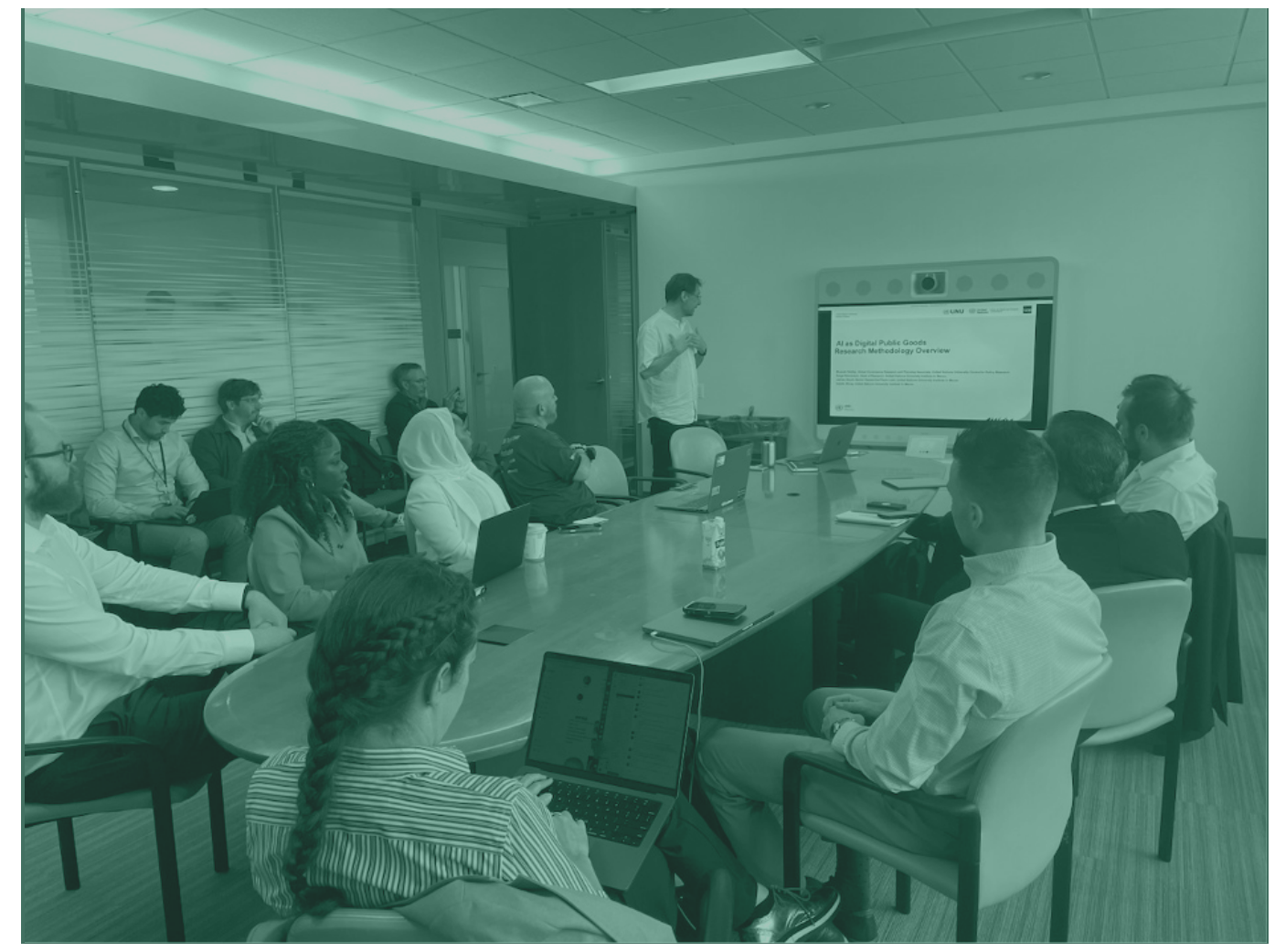

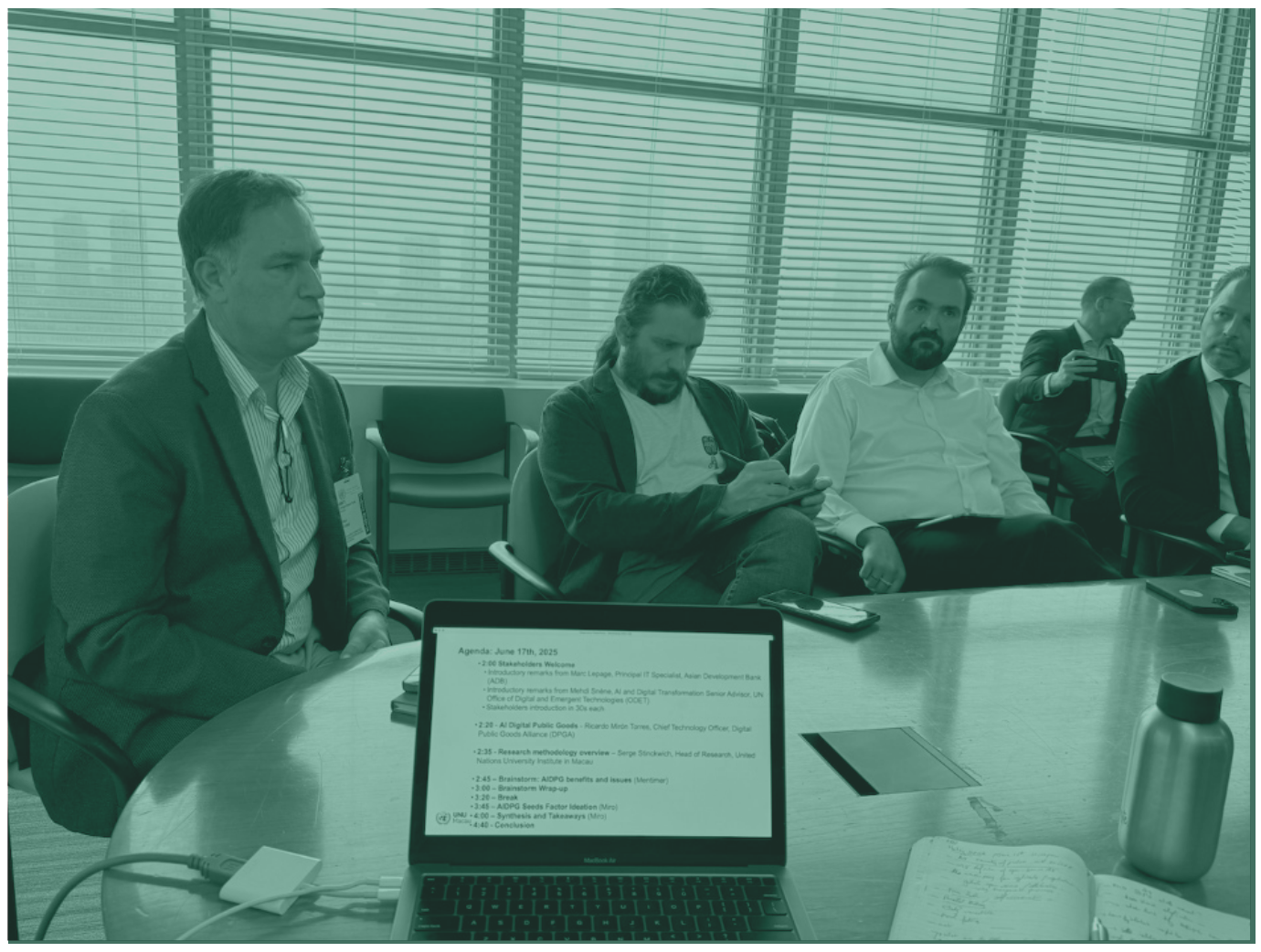

Pictures from Expert Consultation Workshop on Artificial Intelligence as Public Goods

# Section 4: Governance of AIDPGs

This section examines how AIDPGs should be governed across actors and the AI lifecycle. It draws on three complementary evidence streams. A literature review (Section 4.1) situates AIDPG governance within established debates on multi-stakeholder governance and lifecycle-based oversight, identifying both the appeal and the well-documented limits of these models. Qualitative insights from stakeholder interviews (Section 4.2) surface how practitioners across developing and developed countries translate these models into operational expectations, including where their preferred entry points diverge. Quantitative findings from the follow-up survey (Section 4.3) then map which actors and processes stakeholders treat as central to effective AIDPG governance, and where significant gaps remain. Taken together, the three streams point to governance as a continuous, distributed practice rather than a static attribute of an artefact, with implications for the standards, accountability mechanisms, and institutional arrangements developed in the recommendations that follow.

## 4.1 Literature review

### GOVERNANCE AS A PROCESS RATHER THAN A PRODUCT-LEVEL

Conceiving governance as a process rather than a product-level attribute stems from the observation that AI systems do not remain static artefacts amenable to one-time certification: their behaviour shifts as training data accrues, fine-tuning is applied, deployment contexts change, and downstream uses generate feedback that re-enters the development loop. The Model Openness Framework's insistence on transparency across the comprehensive research and engineering lifecycle (White et al., 2024), the DPGA's own treatment of the DPG Standard as a "living framework" subject to recurring revision, and (Gimpel, 2024) call to embed mandatory risk assessments, red teaming, third-party audits, and public reporting into the assessment regime all coalesce around the same argument: governance attaches to the coordinated actions, instruments, and norms through which a community of actors produces, maintains, and adapts a shared resource over time (Tarkowski, 2025), not to a single release version. Reading governance as a process also recasts the assessor's role, since registry inclusion becomes a snapshot within an ongoing relationship of accountability rather than a terminal verdict, as the DPGA's recent introduction of maturity metrics and post-listing review mechanisms acknowledges (DPGA, 2025).

### MULTI-STAKEHOLDER GOVERNANCE

A multi-stakeholder governance is usually regarded as superior for governing complex socio-technical systems (STS), as it strengthens legitimacy and accountability through the inclusion of multiple stakeholder perspectives. DPGA itself operates as a multi-stakeholder initiative, and its CoP constitutes its primary mechanism for norm development and knowledge exchange (Gimpel, 2024). At a broader level, multi-stakeholder governance has deep roots in existing governance frameworks like the Internet Governance, where it has been theorized as a means of pooling diverse expertise and enhancing the legitimacy of rule-making in domains that cut across national jurisdictions. Multi-stakeholder governance is necessary but sometimes not enough. (Gurumurthy et al., 2022) observe that DPGA has not developed binding guidelines on the duties, rights, and obligations of the various actors within DPG partnership ecosystems, relying instead on the assumption that soft norm development in multi-stakeholder communities will safeguard human rights and public interest. The case of the Modular Open Source Identity Platform (MOSIP)[50] (identified as a DPG by DPGA) illustrates how such arrangements might be insufficient: privatized governance within a nominally multi-stakeholder community may lack the enforcement capacity to protect against downstream harms. This critique is consistent with other authors warning against "participation-washing", in which multi-stakeholder processes create an appearance of inclusivity without genuinely redistributing decision-making power (Sloane et al., 2022). A persistent tension exists within public agencies, as the primary beneficiaries of DPG deployment may simultaneously be required to relinquish a degree of sovereign control over the very platforms they depend upon.

(Zhang & Zhang, 2025) conceptualize Large Language Models as quasi-public goods, exhibiting non-rivalry, partial excludability, and significant positive externalities and propose governance structures that optimize social welfare through public-private partnerships.

### LAYERED GOVERNANCE THROUGH THE AI LIFECYCLE

The academic literature converges on the need for governance to be differentiated across the different stages of the AI lifecycle rather than applied uniformly. The OECD's AI system lifecycle framework distinguishes stages of (i) design, data and modelling, including planning and design, data collection and processing, model building and interpretation, (ii) verification and validation, (iii) deployment, and (iv) operation and monitoring, each of which involves distinct actors and generates distinct accountability requirements (OECD, 2019). The AI as DPG Community of Practice Discussion Paper (UNICEF, 2023) similarly maps openness and governance requirements across the components of an AI system (data, code and model) noting that compliance requirements for stand-alone AI models may differ substantially from those for complex, integrated AI solutions.

However, lifecycle governance also needs to be linked to the institutional and regional contexts in which AI systems are deployed. Regional policy frameworks increasingly point to the importance of context-sensitive governance. The African Union's Continental AI Strategy links AI to Africa-centric development priorities and continental ownership (African Union, 2024), the Association of Southeast Asian Nations' governance guidance emphasizes interoperability and context-sensitive implementation (ASEAN Guide on AI Governance and Ethics, 2024), and recent World Bank work places local data, local talent, and public-sector capability at the centre of meaningful AI uptake (Digital Progress and Trends Report 2025, 2025). These frameworks suggest that AIDPGs become politically and institutionally viable only when openness is matched by local capacity to govern data, adapt systems, and retain control over public deployment through institutions capable of oversight, accountability, and course correction.

The governance literature coalesces around several propositions:

(a) Governance of AIDPGs must be conceived as a process rather than a product-level attribute, requiring institutional arrangements that evolve across jurisdictional boundaries.

(b) Multi-stakeholder models, while indispensable, carry well-documented risks of power asymmetry, enforcement deficit, and performative participation; these must be addressed through binding commitments and differentiated role allocation.

(c) Governance should also be distinguished across the whole AI lifecycle.

## 4.2 Qualitative Insights: Stakeholder Interviews on Governance Models

Based on stakeholder insights, the governance of AIDPGs should be approached through two complementary lenses. First, governance must be considered across different levels of stakeholders via a multi-stakeholder, distributed model, emphasizing how international, national, and local actors coordinate. Second, governance should be applied across the AI lifecycle via a modular governance approach, tailoring oversight to the specific phases of data gathering, model production, and downstream implementation.

### MULTI-STAKEHOLDER AND DISTRIBUTED GOVERNANCE: BRIDGING GLOBAL AND LOCAL NEEDS

Interviewees across all regions converged on the need for a multi-stakeholder, distributed governance model to ensure accountability, transparency, and public trust. Trust is viewed not merely as a public sentiment, but as a verifiable attribute achieved through independent assurance, data provenance, and visible public value (Interviewees 9, 10 and 16).

A broad consensus emerged that governance should be "layered"—internationally coordinated yet nationally tailored to local realities. However, a primary divergence emerged regarding how governance should be implemented, largely split between developed countries and developing countries perspectives:

- **Developing countries Perspective (Top-Down Anchoring with Bottom-Up Iteration):** Stakeholders in the developing countries emphasized the need for UN-led coordination and international standard-setting to counter "AI imbalances" and avoid regulatory fragmentation (Interviewees 1 and 2). Due to capacity and resource constraints, these stakeholders prefer strong, top-level guidance that can be adapted locally, preventing the costly reinvention of regulatory frameworks. Furthermore, they advocate for a gradualist approach: setting minimum baselines (such as "negative lists" for prohibited uses) and tightening rules only as concrete risks emerge, ensuring that early, heavy-handed regulations do not stifle beneficial innovation (Interviewees 7, 9, 10 and 16).

- **Developed countries Perspective (Bottom-Up Initiation with Top-Down Consolidation):** Conversely, developed countries interviewees emphasized subsidiarity and local experimentation. Drawing on their more established regulatory capacities, they advocated starting governance at local or municipal levels to surface needs early, enabling faster, context-specific adjustments before wider codifications. These insights can then inform international standards (Interviewees 8, 13 and 20). This approach prioritizes risk-based oversight, scaling regulatory obligations based on the sensitivity of the specific use case (Interviewee 4).

Despite these different entry points, both regions support a layered, bidirectional model (see Figure 7).

**Figure 7:** Layered Bidirectional Model of AIDPG Governance

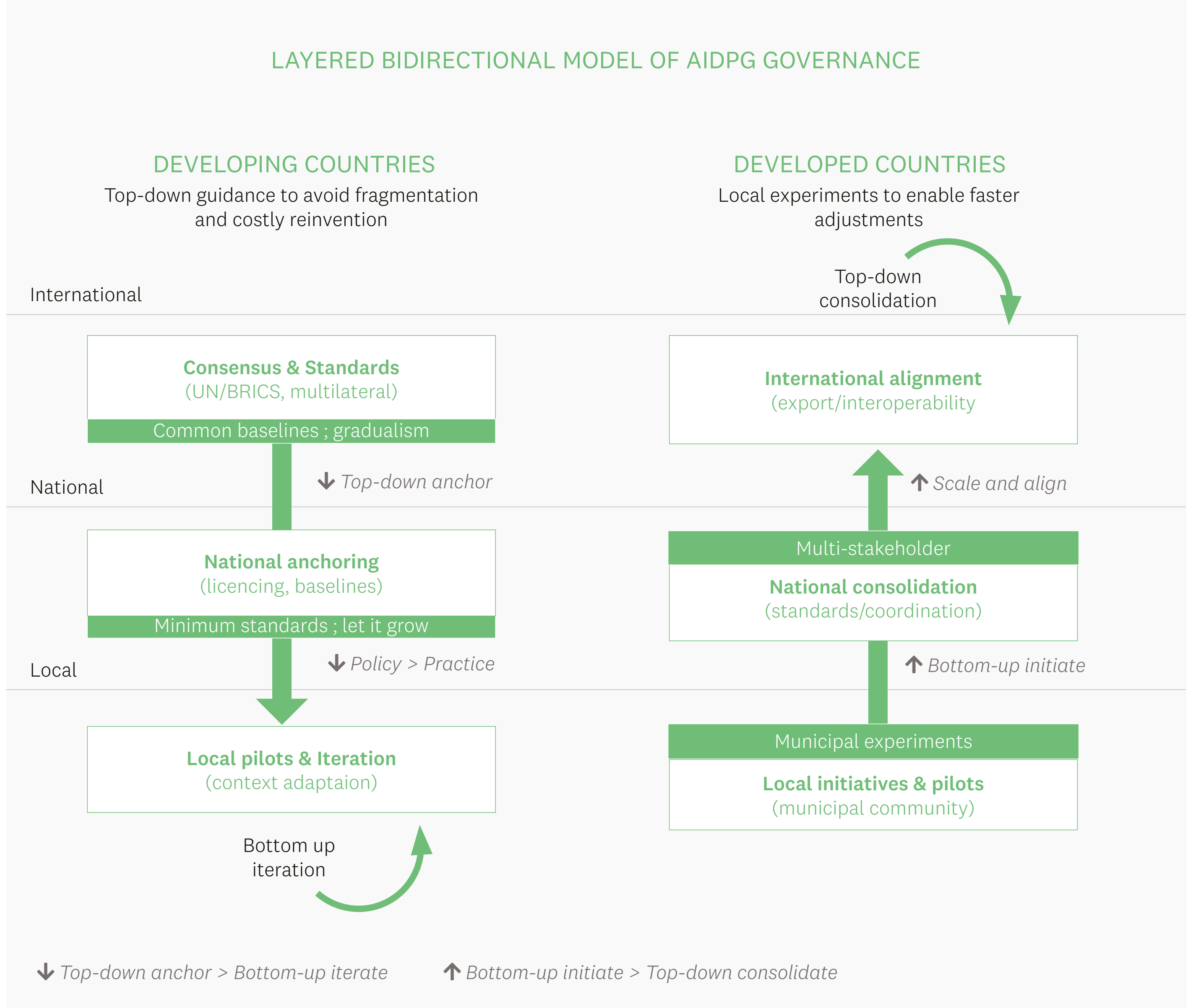

## MODULAR GOVERNANCE ACROSS THE AI LIFECYCLE

In addition to distributing governance across stakeholders, interviewees stressed that oversight must be modular, tailored to each phase of the AI lifecycle to address the non-deterministic nature of AI that sets it apart from traditional software DPGs. This begins with upstream data choices involving consent, data provenance, and preference signalling, which then proceeds through rigorous evaluation and documentation during model development, and culminates in deployment accountability and incident management for downstream impacts. Figure 8 outlines for each phase: the aim and focus, the governance strategies for AIDPGs, and the responsible groups identified by interviewees.

- **Phase 1: Data Stage (Data governance)**
Focuses on upstream choices, such as ethical sourcing, preference signalling, and data provenance. Interviewees emphasized that these data decisions should be strictly governed, as they shape downstream model behaviours.

- **Phase 2: Model Development and Release (Pre-release assurance)**
Focuses on demonstrating the model's adequacy, safety, and transparency before it reaches the public. This phase requires in-house controls, evaluation, third-party audition for pre-trained models, and the publication of artifacts to ensure transparency.

- **Phase 3: Deployment and Implementation (Operational accountability)**
Focuses on post-release oversight to monitor real-world behaviour. Governance at this stage shifts towards managing live incidents monitoring performance drift, ensuring "human-in-the-loop" oversight, and maintaining mechanisms for incident disclosure and redress.

**Figure 8:** AIDPG Modular Governance

### DATA STAGE

*Data governance*

**Focus:** Upstream data choices (consent, provenance, signaling) shape downstream model behavior

| Strategy | Responsible Group(s) |
|---|---|
| **Preference signaling** | Data producers / publishers |
| **Data provenance / labeling** | Data producers ; registries |
| **Ethical sourcing** | Data producers ; Policy/ governance team |

### MODEL DEVELOPMENT & RELEASE

*Pre-release assurance*

**Focus:** Demonstrate adequacy through evaluation and documentation before release

| Strategy | Responsible Group(s) |
|---|---|
| **Versioning & accountable roles**<br>Decide team structure, roles, accountability | Responsible AI / Gov. lead ; Model devs |
| **In-house controls**<br>On data provenance, pipelines, infrastructure | Model developers ; infra/ security teams |
| **Evaluation & testing**<br>eg: invident registers, benchmarking | Model developers ; independant auditors |
| **Auditing pre-trained models**<br>Vendors attestations ; 3rd party audits | Vendors / platforms ; independant auditors |
| **Artifacts Transparency**<br>eg: model / data cards | Model developers / Maintainers |

### DEPLOYMENT & IMPLEMENTATION

*Operational accountability*

**Focus:** Monitor real-world behavior ; manage incidents ; keep humans in the loop ; inform users

| Strategy | Responsible Group(s) |
|---|---|
| **Human-in-the-loop**<br>Usage boundaries / access control | Implementing agencies; product owners |
| **Watermarking / fingerprinting**<br>Enable provenance cues | Platform providers ; standards bodies |
| **Monitoring & drift detection**<br>Performance, safety, bias, drift, misuse | Operators ; Governance teams |
| **Incident registers, discolosure & redress**<br>Operate incident log, intake channels, remediation and communication flows | Oversight bodies ; Implementing agencies / operators (incident response) |

*Notes: Subscripts indicate sources, namely 1 = Interviewee 11 (Developed countries); 2 = Interviewees 9 and 10 (Developing countries); 3 = Interviewee 7 (Developing countries); 4 = Interviewee 20 (Developed countries); 5 = Interviewee 18 (works across developing and developed countries); 6 = Interviewee 5 (Developed countries); 7 = Interviewee 4 (Developed countries).*

*Abbreviations: Gov., Governance; Devs, developers.*

To reduce costs and make this modular governance practical (especially for non-profits and resource-constrained teams in the developing countries) interviewees highlighted two critical strategies:

• **Adapting Existing Enterprise Frameworks:** Rather than building new governance structures ex nihilo, organizations should adapt their existing risk, compliance, and security controls by adding AI-specific checkpoints. This improves adoption feasibility while keeping requirements proportionate (Interviewees 11, 17 and 18).

• **Differentiating between Pre-trained and In-House Models:** Frameworks must distinguish between organizations consuming pre-trained commercial services and those training models in-house. When using pre-trained services, which is a common reality in the developing countries, governance can rely more heavily on vendor attestations and third-party audits. In contrast, in-house training models require deep, resource-intensive governance of data provenance, training pipelines, and infrastructure security (Interviewee 18). Recognizing this distinction ensures that scarce oversight resources are focused where they can have the greatest impact (Interviewees 17 and 18).

The interviewees highlighted that, whether coordinating across different stakeholders or managing the AI lifecycle, AIDPG governance must balance robust risk mitigation with the flexibility required to foster equitable innovation beneficial to the public.
4.3 Quantitative Insights: Drivers and Gaps in AIDPG Governance

To build upon these qualitative insights, a follow-up survey was conducted to identify the central components that ensure the governance processes of AIDPGs are effective in practice. These specific components were selected for the survey because they emerged as the most frequently mentioned and highly prioritized elements during the interview discussions on AIDPG governance. To examine how key actors and governance processes interact to regulate AIDPGs, we mapped our survey results using a force-directed correlation network (see Figure 9). In this visual map, participants rated the importance of various actors and processes on a 5-point scale (ranging from 1= not important at all to 5 = extremely important). Items with highly correlated ratings are drawn closer together and connected by thicker lines. Consequently, a component located at the centre with many connections represents a core hub of the governance system.

As illustrated in Figure 9, the most prominent finding is that citizen representation occupies a central position in AIDPG governance. It acts as the core hub, linking all the key actors and processes identified by our stakeholders. This quantitative finding reinforces that citizen involvement is central to ensuring AIDPGs accurately address public needs and remain safe and beneficial.

**Figure 9 (a and b):** Force-directed correlation network presenting the strength of associations among key actors and governance processes of AIDPGs

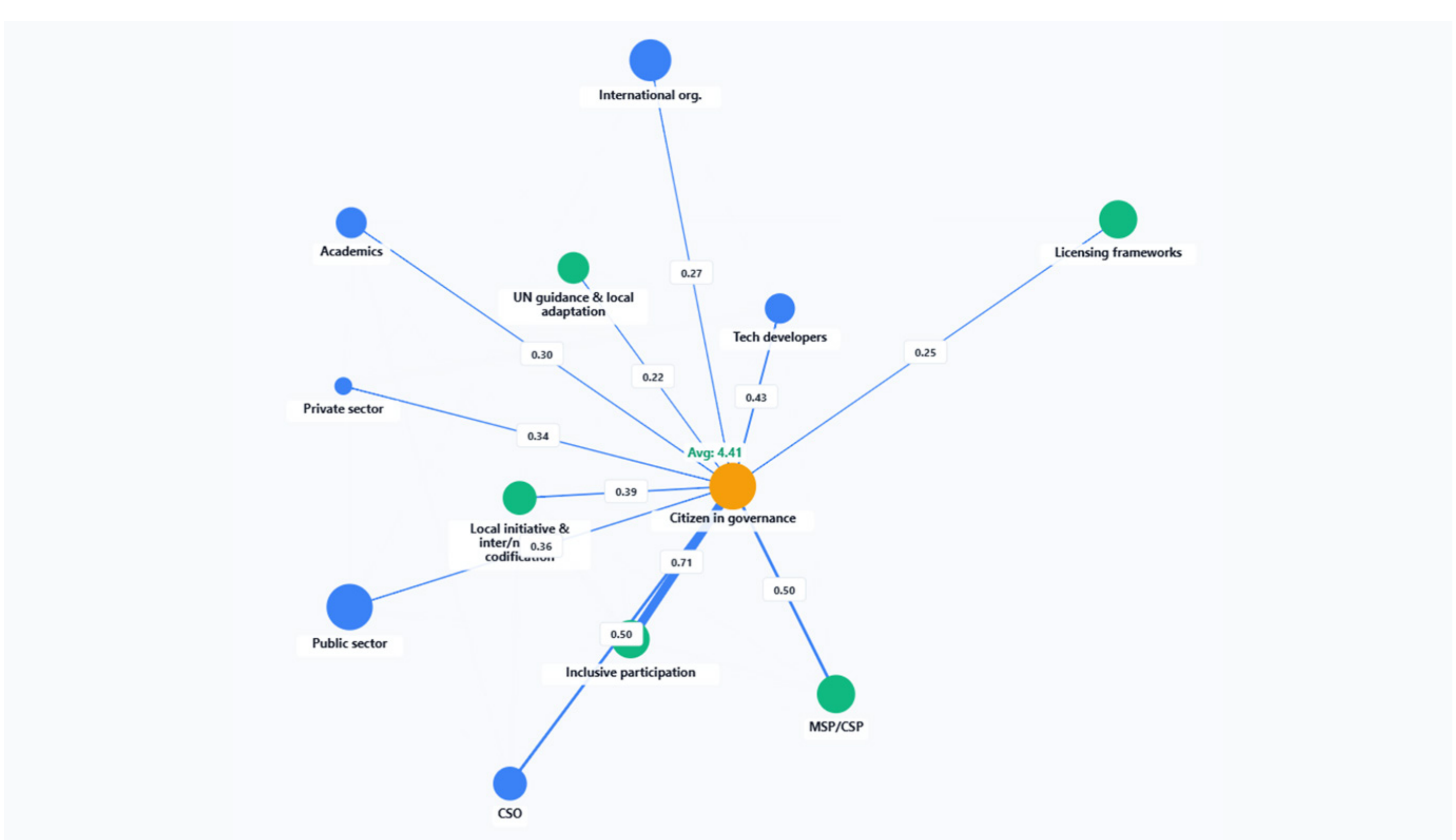

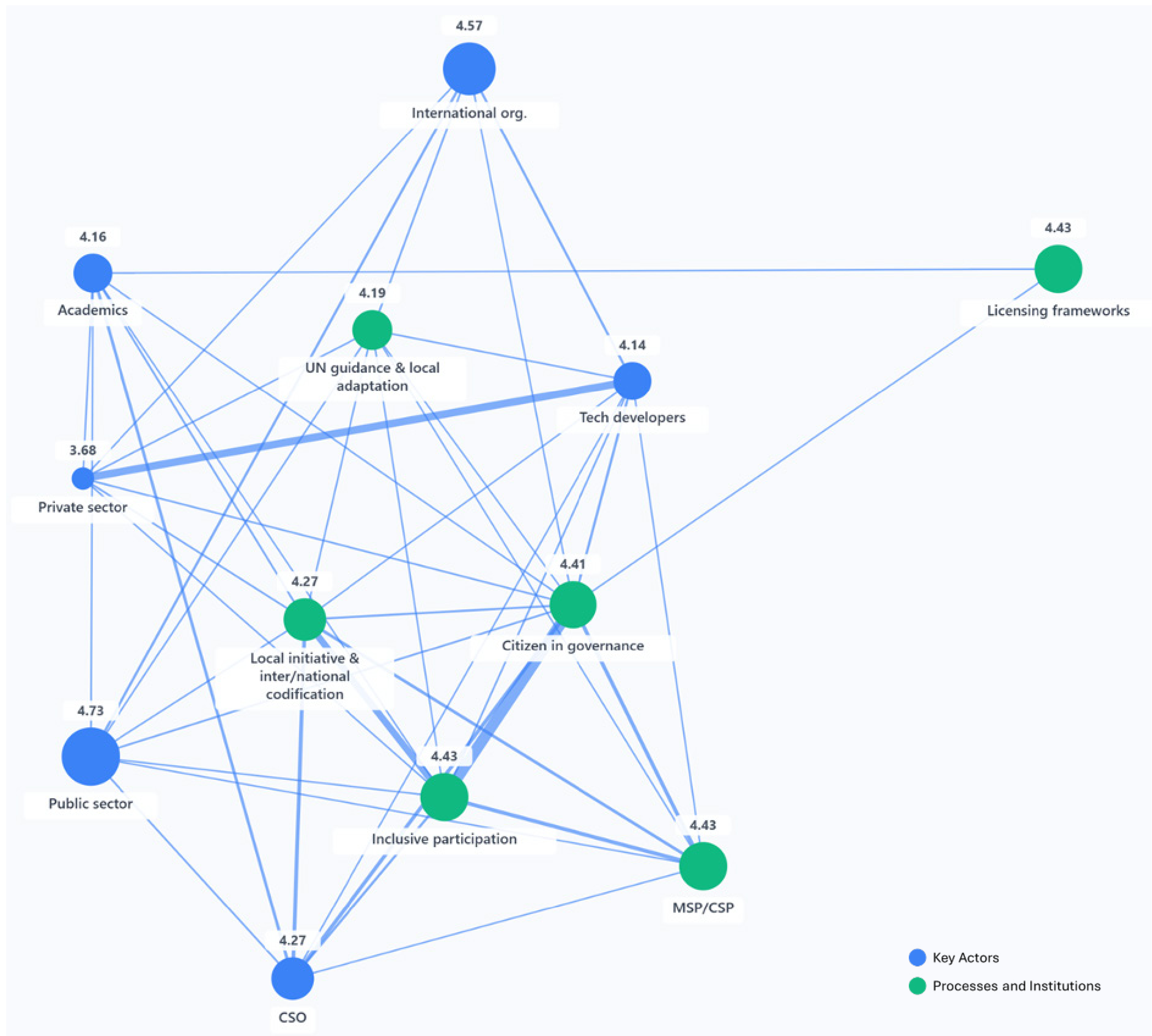

*Note: The node size corresponds to the average rating of each element, while the thickness of the connecting blue lines denotes the strength of their association. The number on top of each node represents its average rating, and the number overlaying each line indicates the correlation coefficient. Figure 9a presents the complete force-correlation network and Figure 9b presents the elements citizen in governance is associated with and the strength of its association.*

*Abbreviations: MSP/CSP, Multi-stakeholders collaboration/Cross-sector collaboration.*

Examining the associations between the multi-stakeholder distributed governance approaches, raised by developing countries and developed countries stakeholders during the interviews, and the different key actors in the system further illuminates the institutional drivers behind these two pathways. We observed a clear split between who leads these respective governance processes. The bottom-up approach (which moves from local initiatives to international or national codification) is uniquely associated with Civil Society Organizations (CSOs) as seen in Figure 9. Conversely, the top-down approach (starting with UN guidance and moving to local adaptation) is uniquely tied to international organizations. These results highlight how different key actors are perceived to be associated with distinct governance pathways.

Lastly, the survey revealed an important gap regarding licensing. While licensing frameworks have been raised as a vital part of openness and the governance process, our analysis shows that they are associated with one key actor (academics) and one process (citizen governance). This highlights that although licensing is an important mechanism, further work is needed to ensure its comprehensive integration into the roles of all key actors and governance processes for AIDPGs.

## 4.4 Conclusion

As AIDPGs are increasingly understood as socio-technical ecosystems rather than stand-alone artefacts, their governance remains insufficiently specified in current policy and standards discussions. While the DPG Standard provides a basis for alignment with relevant standards and best practices through Indicator 8, the practical question is how to operationalize and govern such alignment for AI systems. The support for and governance of this alignment becomes necessary so that stakeholders can compare AIDPGs with other forms of AI systems and determine which systems enter public service, based on what evidence, under which safeguards, and with what scope to adapt, contest, or discontinue them over time.

The findings in this section show that AIDPG governance cannot rely on a single model. Stakeholders support layered governance, but differ on where it should begin. Actors operating with greater dependence on foreign models, infrastructure, and assurance mechanisms will likely prioritize international baselines, localization, sovereign capacity, and protections against lock-in. Actors with stronger domestic technical and regulatory capacity tend to emphasize subsidiarity, local experimentation, and risk-based adaptation.

Governments and public institutions are more likely to invest in AIDPGs when they can govern them as public infrastructure rather than merely consuming them as external tools. A workable governance approach, therefore, needs to combine common baseline requirements with context-sensitive implementation. For public institutions, localization must be understood as a governance condition, not only a technical adjustment. This includes the ability to align models and data practices with local law, language, administrative practice, procurement rules, and public priorities; to test systems against local harms and performance expectations; and to maintain, adapt, or replace systems without long-term lock-in. Data governance is central to this task, especially where AIDPGs rely on sensitive, administrative, or community-linked data.

AIDPG governance also requires a clearer allocation of responsibility across the value chain. Developers shape upstream choices, such as design, training, documentation, and data provenance, while deployers determine how systems enter institutions, what safeguards apply, and how redress operates in practice. Users are central to identifying context-specific failures and judging whether systems generate public value in practice. The governance challenge arises from a misalignment of responsibilities across the AI lifecycle: developers shape systems they may not later control, deployers govern systems they may not fully understand, and users may lack formal channels to contest harms. These gaps become sharper when systems are adapted, integrated into services, or reused in multiple contexts.

# Section 5: Conclusion

This report argues that AI systems can function credibly as Digital Public Goods (DPGs) only if they are assessed not merely as open artefacts, but as socio-technical systems whose public value depends on data governance, contextual deployment, safeguards, accountability, and long-term stewardship.

The report contributes to this debate by shifting attention from narrow questions of AI openness towards a system-level governance and assessment framework. Existing DPG and open source AI frameworks provide important foundations, particularly in terms of licensing, transparency, and access. However, they remain insufficient to address the distinctive features of AI systems, including opaque data provenance, uneven deployment conditions, downstream harms, and distributed responsibility among developers, deployers, users, and institutions.

The findings also show that openness and public value are not equivalent. Stakeholders value accessibility, transparency, SDG alignment, privacy, and responsible AI safeguards, but these attributes do not always point in the same direction. AIDPG assessment must therefore distinguish between systems that are merely open, technically reusable systems, and systems that can demonstrate public value under specific deployment conditions. The report also shows that governance preferences differ according to actors' positions in the AI ecosystem: those more dependent on external models, infrastructure, and assurance mechanisms place greater emphasis on international baselines, localization, sovereign capacity, and protection against lock-in.

A key finding is that AI systems may be labelled or treated as public goods before they have demonstrated clear public value, contextual adequacy, accountability, or sustainability. This creates a risk of premature public-good claims, where openness or stated SDG alignment substitutes for evidence that the system is safe, locally appropriate, maintainable, and beneficial in practice. For AIDPGs to serve public institutions, particularly in lower-capacity contexts, recognition frameworks therefore need to be supported by clearer evidence requirements, accountability mechanisms, financing arrangements, and localization capacities.

In terms of measurement, AIDPGs do not necessarily require a wholly separate measurement universe, but they do require a clearer bridge to connect any future AIDPGs to existing sustainability and ESG assessment and reporting approaches. Openness and declared SDG alignment are not sufficient to demonstrate public value, since they do not show who benefits, what risks are managed, or whether the system remains sustainable over time. Therefore, the main gap is the lack of a clear mechanism to link AI-system assessment to adjacent, evolving impact domains that offer valuable entry points for meaningful evaluation to benefit all AIDPG stakeholders. Bridging this gap requires both adaptation to AIDPG-specific requirements and a formal mechanism for continuous revision, ideally supported by a designated body tasked with monitoring changes and updating the framework over time.
Two governance questions remain especially open: how responsible data sharing and stewardship should operate when full openness proves infeasible or undesirable, and how responsibilities should be allocated once others deploy, adapt, and reuse AI systems.

The contrast in this report between top-down anchoring and bottom-up experimentation reflects these uneven levels of bargaining power and institutional capacity across the AI stack. Actors that rely more heavily on external models, cloud infrastructure, and third-party assurance need stronger international reference points because they face higher switching costs and weaker negotiating power. Actors with stronger domestic regulatory and technical capacity can rely more on local experimentation and iterative rule-making. AIDPG governance, therefore, needs to address interdependence directly because public-value claims are shaped not only by the openness of an artifact but also by who controls the infrastructures, standards, and decision points on which it depends.

This report shows clear demand for international baselines, but no single UN institution currently provides a comprehensive home for this agenda, and the institutional landscape remains fragmented. Within the current UN AI governance support structure, UN ODET[51] carries the UN's broader digital and AI coordination role, the United Nations Educational, Scientific and Cultural Organization (UNESCO) anchors the global ethics standard (Recommendation on the Ethics of Artificial Intelligence - UNESCO Digital Library, 2021), the International Telecommunication Union (ITU) plays a growing role in standards and technical convening (ITU, 2025), and the United Nations Development Programme (UNDP) remains deeply engaged in the DPG[52] and DPI ecosystem, including through the DPGA (UNDP, 2023). For AIDPG governance, the task is not to locate a single institutional centre, but to connect these functions to collaborative development of some global baseline governance norms. A workable governance architecture for AIDPG would therefore combine internationally legible baseline requirements, regional cooperation on interoperability and shared capacity, and national authority over procurement, localization, deployment, and evaluation.

# Section 6: Strategic Recommendations for AIDPGs

The recommendations outlined in this report are intended as non-binding inputs for the wider DPG ecosystem, including funders and donors, governments, and implementing agencies, as well as for the DPG Standard Council on matters within its remit. Several of the issues raised fall beyond the DPGA's current mandate and call for coordinated action across stakeholders.

The report advances ten recommendations, organized into four thematic clusters captured by the SAFE mnemonic: Standard, Accountability, Finance, and Equity.

- **S - Standard.** What might go into the AIDPG Standard itself: the substantive and procedural criteria against which AI systems should be assessed for DPG certification.
- **A - Accountability.** How the ecosystem is organized. Recommendations in this cluster allocate roles across certifying bodies, developers, deployers, and the communities affected by deployment.
- **F - Finance.** The enabling infrastructure and financing mechanisms required for the ecosystem to operate sustainably, including shared compute, data commons, and long-term stewardship funding.
- **E - Equity.** What has to be in place before adoption in the developing countries can be called equitable: local capacity, and certification criteria that take developing countries deployment contexts seriously.

## 6.1 S. AIDPG Standard Recommendations

### S1. ADOPT THE MODEL OPENNESS FRAMEWORK AS THE REFERENCE GUIDELINE FOR AIDPG OPENNESS ASSESSMENT.

Rather than inventing an ad hoc openness metric, any AIDPG framework should adopt or formally align with the Model Openness Framework (White et al., 2024), which provides a three-tier classification (Open Model, Open Tooling, Open Science) and specifies which components (architecture, weights, training code, datasets, documentation) must be released under which licence types. This would provide a common vocabulary for assessing openness across the AIDPG landscape and would make the DPG Standard more interoperable with the broader AI openness ecosystem. The OECD's AI Openness Primer for Policymakers (OECD, 2025) adopts the MOF as its reference taxonomy, which lends it additional institutional legitimacy.
The purpose of adopting the MOF is not to create a single pass/fail test for openness, but to provide a common vocabulary for describing different degrees and forms of openness across AI systems.

### S2. ESTABLISH A PUBLIC-INTEREST DATA STEWARDSHIP PATHWAY FOR AIDPGS.

Where AIDPGs depend on sensitive, administrative, or community-linked data, policy should move beyond the binary choice between full openness and ad hoc restriction. Governments and standard-setting bodies should create a formal stewardship pathway for such systems, allowing governed access models such as data trusts, data commons, or other institutional access arrangements, provided they specify lawful basis, representation, access conditions, benefit-sharing, review procedures, and institutional responsibility for misuse or mission creep.

### S3. BUILD PUBLIC-SECTOR READINESS EVIDENCE FOR AIDPGS INTO THE AIDPG STANDARD.

The AIDPG Standard should require AI systems intended for public service use to provide a simple evidence package demonstrating their suitability, safety, and usability in the intended context, particularly in the developing countries. This should include basic information on the system's purpose, data sources, known limitations, safeguards, human oversight, redress options, update conditions, local-language testing, and context-specific evaluation. The AIDPG framework should help make this evidence available upfront so that public institutions, especially in lower-capacity settings, do not have to assess each system ex nihilo.

### S4. CREATE AN AIDPG PUBLIC VALUE AND RISK ANNEX WITH A SMALL MANDATORY INDICATOR SET

The AIDPG standard should move beyond eligibility and principle-based declarations by adding a concise annex of measurable indicators that every applicant or deployer can report against. That annex should cover social value, economic value, risk, long-term sustainability, and maintainer engagement and community health (recognizing that many AIDPGs depend on active open-source communities for ongoing adaptation and bug reporting). The purpose is to create a common minimum evidence package that allows governments and other public actors to compare systems beyond openness and stated SDG alignment. This recommendation is distinct from the CoP's existing focus on do-no-harm testing and templates: it addresses the still-unresolved question of how broader public value, trade-offs, and implementation conditions should be measured consistently. Where possible, the annex should align with existing ESG and sustainability reporting frameworks to enable comparable disclosure on social, environmental, governance, and economic impacts, including privacy, supplier social conditions, energy use, and emissions. AI-specific frameworks should remain primary for system-level risk, context, and governance, and corporate ESG disclosure should not be treated as a substitute for product-level public-interest assessment.

## 6.2 A. Accountability and Governance

A1. BUILD A CONSOLIDATED AIDPG GOVERNANCE SUPPORT STRUCTURE AROUND THE EXISTING DPGA ECOSYSTEM.

The wider governance support required for AIDPG development and deployment remains fragmented across institutions, but creates a clear opportunity to extend the DPGA's role beyond recognition and standard setting. Important elements of this support are already emerging through the work of DPGA members, including UN ODET, UNESCO, and ITU, as well as through regional initiatives. DPGA should remain the anchor for recognition, standards, and cross-case learning, while linked institutional efforts should be brought into a more coherent AIDPG governance guidance support structure that connects normative guidance, technical standards, interoperability work, implementation support, and public-sector adoption. This would help close the governance gap without creating a parallel institutional architecture.

A2. REQUIRE RESPONSIBILITY MAPS AND REDRESS MECHANISMS ACROSS THE AIDPG VALUE CHAIN.

AI systems classified as DPG should carry a published responsibility map that distinguishes the obligations of developers, deployers, procuring authorities, maintainers, and auditors. At a minimum, this should allocate responsibility for data provenance, documentation, model updates, incident response, user notification, audit access, complaint handling, and remedy. For higher-impact systems, this should be paired with formal channels for user contestation and public-interest oversight.

## 6.3 F. Finance and enabling Infrastructure

F1. DESIGN A PUBLIC-INTEREST COMPUTE ACCESS STRATEGY FOR AIDPG DEVELOPMENT, FINE-TUNING, AND INFERENCE.

The significant capital investment required for developing AI poses a structural barrier for participation, particularly among institutions in the developing countries. Training a single large language model can cost $300–400 million, while existing computational resources are often insufficient and restricted to well-resourced academia in high-income countries. Open source AI cannot deliver on its promise without public infrastructure for inference and post-training. An AI-DPG framework should therefore recommend: (a) the establishment or expansion of public compute facilities accessible to AI-DPG developers and deployers, drawing on pooled-compute models such as the US National AI Research Resource (NAIRR), the EU's High Performance Computing Joint Undertaking (EuroHPC JU), Japan's AI Bridging Cloud Infrastructure (ABCI), and new initiatives such as the ICTP Consortium for Scientific Computing (ICOMP) and the International Computation and AI Network (ICAIN), with mandates that explicitly serve developing countries needs; (b) multilateral development bank and donor co-financing of compute access for low- and middle-income country (LMIC) institutions engaged in AI-DPG adaptation and deployment; (c) tiered access models that prioritize public-interest use cases, with transparent eligibility criteria and independent review; and (d) governance arrangements — including open standards, portable workloads, and contractual safeguards on data and model weights that prevent compute facilities from becoming vectors for vendor lock-in or data extraction. Sustained public investment in shared compute capacity is essential to counteract the AI compute divide and to reduce societal dependence on Big Tech infrastructure.

F2. LINK FUNDING, PROCUREMENT, AND RENEWAL DECISIONS TO PERIODIC REPORTING ON OUTCOMES, NOT ONLY TO THE INITIAL AIDPG DESIGNATION.

If measurable standards are meant to influence real adoption decisions, they need to be tied to procurement and financing practices. UN agencies, MDBs, and national governments should require a short ex ante and periodic ex post evidence package for AIDPGs used in public programmes, covering service reach, context fit, reuse and adaptation, cost implications, incidents or harms, maintenance arrangements, and energy or resource use where relevant. This would shift the AIDPG Standard from a one-time recognition model towards an evidence-based public-value model, making the designation more decision-useful for public actors.

## 6.4 E. Equity in the developing countries

E1. SUPPORT THE DEVELOPMENT OF LOCAL LANGUAGE AND DOMAIN-SPECIFIC TRAINING DATA AS A PUBLIC-GOOD INFRASTRUCTURE.

Many AIDPGs will underperform or fail entirely in deployment contexts where local language data is scarce or where domain-specific data (health records, agricultural extension, legal corpora) has not been curated for AI use. Capacity building should include: (a) public investment in local-language data collection and annotation, treated as infrastructure rather than project expenditure; (b) governance frameworks for community-linked data that respect Indigenous data sovereignty principles; (c) open data repositories with appropriate licensing for AIDPG training and fine-tuning; and (d) partnerships between national statistical offices, universities, and local AIDPG developers to produce high-quality, contextually grounded datasets.

E2. INVEST IN AIDPG'S LOCAL EVALUATION AND AUDIT CAPABILITIES AS A PREREQUISITE FOR AIDPG ADOPTION.

Governments in lower-capacity settings cannot exercise meaningful oversight over AIDPGs if they lack the institutional capability to evaluate, test, and audit AI systems. Capacity building recommendations should include: (a) support for national or regional AI evaluation centres that can conduct context-specific testing (bias, fairness, language adequacy, domain accuracy); (b) training programmes for public-sector procurement officers and IT staff on AI system evaluation, drawing on the deployment assurance profile; (c) South–South knowledge exchange networks for sharing evaluation tools, test datasets, and deployment lessons; and (d) university-

**Figure 10:** SAFE recommendations

| S: STANDARD<br>*Certificate criteria* | A: ACCOUNTABILITY<br>*Roles, redress, oversight* |
|---|---|
| **S1:** Model Openness Framework<br>**S2:** Data stewardship pathway<br>**S3:** Deployment assurance profile<br>**S4:** Public-value indicator annex | **A1:** Governance support structure<br>**A2:** Responsability maps and redress mechanism |
| **F: FINANCE**<br>*Enabling infrastructure* | **E: EQUITY**<br>*Developing countries Adoption* |
| **F1:** Public-interest compute access<br>**F2:** Outcome-linked funding | **E1:** Local language and domain data<br>**E2:** Local evaluation capability |

# Appendices

## Appendix A. Detailed Literature Review Methodology

This study used a structured comparative desk review to map the evolving literature and policy debate on Artificial Intelligence as Digital Public Goods (AIDPGs), Digital Public Goods (DPGs), and open source AI. The review aimed to clarify key concepts, identify major areas of agreement and contestation, and provide an evidence base for the subsequent interview and survey phases. It was therefore intended as a policy-oriented scoping exercise rather than a full systematic review.

The review focused on five thematic areas defined at the outset of the project: definitions and governance; licensing and openness; benefits; risks and development implications. Searches were conducted across both academic and institutional sources. Academic searches covered Google Scholar, JSTOR, IEEE, and Scopus, while institutional and policy sources included the ADB, DPGA Registry, World Bank, OECD, and the European AI Office. The review also drew on relevant case studies from development and AI initiatives, particularly those that could illustrate practical implementation issues in low- and middle-income countries.

Combinations of keywords relating to AI public goods, digital public goods, open source AI, public interest, and AI for social good guided the searches. Only sources that substantively contributed to definitional debates, openness and licensing issues, governance questions, development implications, or practical examples relevant to AIDPGs were included. Duplicates and sources with limited relevance were discarded. The summarized and tagged sources were incorporated into a shared workflow to facilitate themes comparison and reduce duplication in the synthesis.

The findings were synthesized thematically through iterative team discussion. This process helped identify areas of convergence, divergence, and unresolved debate, particularly around definitions, the meaning of openness across different AI components, data governance, and the relationship between public-interest goals and practical implementation constraints. The desk review also informed the development of the interview guide and survey instrument by identifying the main issues requiring further multi-stakeholder input. As such, the literature review formed the conceptual foundation of the report's broader mixed-methods design.

## Appendix B. Interview Methodology

KEY INFORMANT INTERVIEWS

The semi-structured interview captured experts' perspectives on AIDPGs while situating responses within their professional and geographic backgrounds. After obtaining informed verbal consent, interviews began with questions on the interviewee's role, sectoral affiliation (e.g. policymaking, academia, civil society, private sector), and geographic location. This information was important for interpreting how sectoral priorities and regional contexts, including perspectives from both the developed countries and developing countries, might shape views on AIDPGs. The interview protocol addressed topics such as definitions, licensing and ethical use, governance and trust, benefits and risks, sustainability, and issues of access and equity. Open-ended questions, supported by optional prompts, ensured critical topics were addressed while allowing participants to highlight areas of greatest relevance to their expertise. Each interview concluded with space for final reflections and recommendations of additional experts, thereby ensuring breadth and inclusivity in the perspectives gathered.

## PARTICIPANTS SELECTION

Participants were purposively selected for their expertise and leadership in the field of AIDPGs. Invitations to participate in semi-structured interviews were extended to individuals recognized as knowledge experts and thought leaders, with deliberate inclusion of representatives from both the developed countries and the developing countries to ensure geographic and contextual balance.

Potential participants were identified through three main channels:

1. Networks of experts working on AIDPGs;

2. Referrals from the Asian Development Bank (ADB) or United Nations Office for Digital and Emerging Technologies (UN ODET), based on prior collaborations and professional networks;

3. Individuals who had previously expressed willingness to be contacted for this research project.

Selection criteria emphasized expertise and thought leadership across key sectors, including policymaking (UN agencies, international organizations, and government representatives), private sector and technical development (open source AI and DPG practitioners), civil society, and academia.

## INTERVIEW PROCEDURE

Interviews were held either in person at the meeting venues with participants attending the ITU AI for Good Global Summit 2025[53] in Geneva and the AIxADB[54] 2025 “Harnessing AI for Public Sector Innovation and Fiscal Resilience” in Singapore, or online for those not attending or unable to make time for an in-person interview during the summit. Informed consent was obtained from the participants prior to the interview. The interviews were conducted in English and lasted 30–60 minutes. Only the audio of the interviews was recorded.

## INTERVIEW PARTICIPANTS

The final sample consisted of 21 knowledge experts and thought leaders on AIDPGs, representing a range of sectors including policymaking (UN agencies, international organizations, and government bodies), private sector (start-ups and large tech companies) and technical development (open source AI and DPG practitioners), civil society, and academia. Participants were drawn from both the developed countries and the developing countries, ensuring a balanced representation of perspectives across regions and domains of expertise. Table 6 provides an overview of the interview sample composition.

**Table 6:** Interview Sample Composition

| Ref. No. | Sectors | Policy-making | Technical dev. | Academia Research | Civil society | Private Sector | Global |
|---|---|---|---|---|---|---|---|
| 1 | Private sector | | | | | √ | Developing countries |
| 2 | Private sector | | | | | √ | Developing countries |

| Ref. No. | Sectors | Policy-making | Technical dev. | Academia Research | Civil society | Private Sector | Global |
|---|---|---|---|---|---|---|---|
| 3 | Policymaking and civil society | √ | | | √ | | Developed countries |
| 4 | Policymaking and technical development | √ | √ | | | | Developed countries |
| 5 | Technical development | | √ | | | | Developed countries |
| 6 | Academia | | | √ | | | Developed countries |
| 7 | Policymaking/UN | √ | | | | | Developing countries |
| 8 | Technical development/ research | | √ | √ | | | Developed countries |
| 9 | Policymaking | √ | | | | | Developing countries |
| 10 | Private sector | | | | | √ | Developing countries |
| 11 | Policymaking and technical development | √ | √ | | | | Developed countries |
| 12 | Policymaking/ research | √ | | √ | | | Developed countries |
| 13 | Technical development and civil society | | √ | | √ | | Developed countries |
| 14 | Academia and civil society | | | √ | √ | | Developed countries |
| 15 | Policymaking and government | √ | | | | | Developing countries |
| 16 | Policymaking and government | √ | | | | | Developing countries |
| 17 | Technical development (ADB) | | √ | | | | Developed countries/ Developing countries |
| 18 | Technical development (ADB) | | √ | | | | Developed countries/ Developing countries |
| 19 | Private sector (Large tech) | | √ | | | √ | Developed countries |
| 20 | Private sector (Large tech) | | √ | | | √ | Developed countries |

**Supplemental Table.** Ratings of AIDPGs core attributes across different regions

| Attribute | Definition | Whole Group Mean (SD) | Developed countries Mean (SD) | Developing countries Mean (SD) | Higher AI Readiness Mean (SD) | Lower AI Readiness Mean (SD) |
|---|---|---|---|---|---|---|
| Free of charge | 1. Free of charge at the point of access | 4.03 (1.12) | 3.73 (1.49) | 4.23 (0.75) | 3.79 (1.25) | 4.46 (0.66) |
| Freely accessible | 2. Freely accessible (without third-party interference) | 4.70 (0.66) | 4.80 (0.56) | 4.64 (0.73) | 4.58 (0.78) | 4.92 (0.28) |
| Do no harm | 3. Do no harm | 4.62 (0.72) | 4.47 (0.99) | 4.73 (0.46) | 4.58 (0.83) | 4.69 (0.48) |
| Beneficial for SDGs | 4. Beneficial for the public and help to achieve SDGs | 4.59 (0.60) | 4.47 (0.64) | 4.68 (0.57) | 4.54 (0.59) | 4.69 (0.63) |
| Open-source | 5. Open-source (at least to a certain extent) | 4.38 (0.72) | 4.67 (0.62) | 4.18 (0.73) | 4.46 (0.78) | 4.23 (0.60) |
| Transparent | 6. Transparent (at least to a certain extent) | 4.59 (0.60) | 4.67 (0.62) | 4.55 (0.60) | 4.50 (0.66) | 4.77 (0.44) |
| Inclusive development | 7. Inclusive development/Leaving no one behind | 4.46 (0.90) | 4.20 (1.01) | 4.64 (0.79) | 4.33 (0.92) | 4.69 (0.85) |
| Privacy/Data Safety | 8. Comply with privacy laws and safeguard user data | 4.62 (0.83) | 4.47 (0.99) | 4.73 (0.70) | 4.50 (0.98) | 4.85 (0.38) |
| Responsible AI | 9. Responsible AI attributes (e.g. explainability, fairness, transparency) | 4.54 (0.77) | 4.47 (1.06) | 4.59 (0.50) | 4.50 (0.88) | 4.62 (0.51) |
| Adaptable context | 10. Adaptable for local contexts and adoption | 4.46 (0.87) | 4.33 (1.11) | 4.55 (0.67) | 4.46 (0.93) | 4.46 (0.78) |

*Note: Rating ranges from 1 to 5.*
*Abbreviations: SD, Standard Deviation; SDGs, Sustainable Development Goals.*

## Appendix C. Survey Methodology

KEY INFORMANT SURVEYS

Building upon the rich qualitative data gathered during the semi-structured interviews, a targeted key informant survey was developed to validate and quantify these initial findings. The survey instrument was directly informed by the core themes that emerged from the interviews, specifically focusing on the definition of AIDPGs, preferred governance structures, and the unique opportunities and challenges associated with AIDPG deployment in the developing countries. This phase aimed to assess consensus and divergence across a different pool of stakeholders.

To draw robust quantitative insights from the survey data, advanced analytical tools were employed to confirm and extend the observed associations between different stakeholder perspectives. Notably, force-directed correlation networks were utilized to map the complex relationships between various views and priorities. This analytical approach enables the identification of "central" variables, i.e., highly connected elements that serve as structural anchors within the network. Furthermore, this approach visually and statistically highlights distinct clusters of agreement as well as critical areas of tension. These quantitative insights provide an evidence-based foundation for informing policy, allowing decision-makers to clearly identify key priorities and navigate competing paradigms surrounding the definition, governance, and equitable implementation of AIDPGs.
Participant Recruitment

In January 2026, the research team set out to recruit prospective participants, identified as subject matter experts in AI systems, by disseminating survey links through targeted email invitations and professional networks (LinkedIn). The recruitment materials provided a comprehensive overview of the study's aims and objectives.

DATA COLLECTION PROCEDURE

Data collection was conducted from January to February 2026 using the online survey instrument, LimeSurvey[55], an open-source survey platform self-hosted by the research team. Prior to commencing the survey, explicit informed consent was obtained from all participants.

SURVEY PARTICIPANTS

The final sample consisted of 37 knowledge experts and thought leaders on AIDPGs, representing a range of sectors. These include intergovernmental organizations (such as UN agencies and affiliated programs), the private sector, technical developers, civil society, and academia. Figure 11 provides a detailed breakdown of participant distribution by sector. Notably, the respondents are affiliated with several prominent institutions at the forefront of this field, such as Creative Commons and MIT Computer Science and Artificial Intelligence Laboratory (MIT CSAIL).

**Figure 11:** Survey Participant Distribution by Sector

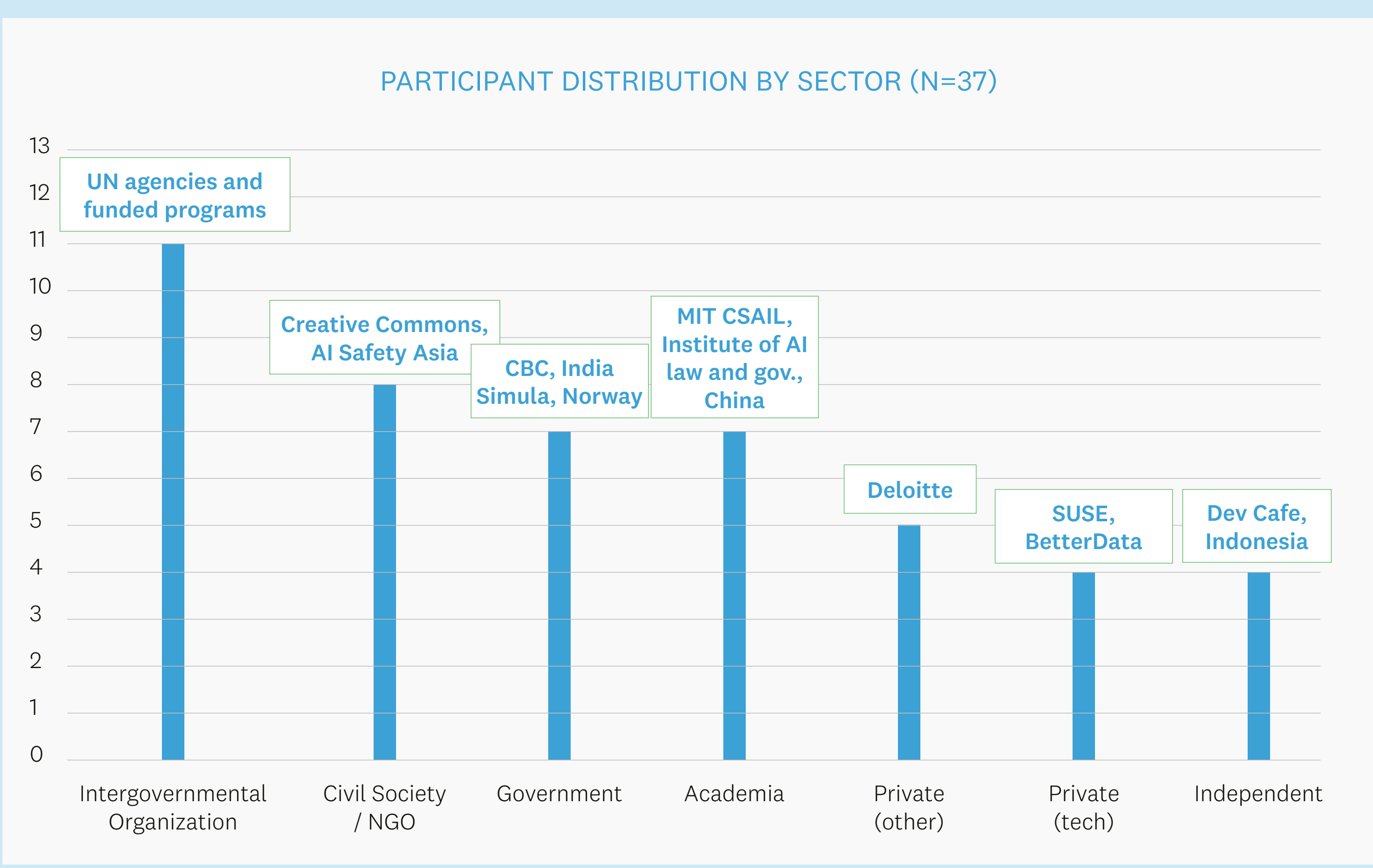

To ensure a diverse representation of global perspectives, the survey targeted participants across a wide spectrum of government AI readiness and national income levels, deliberately encompassing both the developed countries and the developing countries. Consequently, the final sample successfully captures a broad cross-section of regional contexts and technical capacities. Figure 12 provides an overview of the participant distribution according to these economic and AI-readiness indicators.

**Figure 12:** Participant Distribution by Government AI-Readiness and Economic Indicators

AI readiness global rank in brackets
AI Readiness of Participants' Countries
Global Avg
Participant Mean
0
25
50
75
95
AI Readiness Score
High income
Upper middle income
Lower middle income
Low income
Global average = 42.5 (Oxford Insights, 2025)
High Income
United States (1)
France (2)
Netherlands (4)
Germany (6)
Singapore (7)
Norway (10)
Italy (25)
Switzerland (28)
Belgium (30)
Upper Middle Income
China (8)
Brazil (22)
Thailand (34)
Malaysia (38)
Indonesia (46)
Argentina (56)
Jamaica (99)
Lower Middle/ Low Income
India (27)
Philippines (43)
Uzbekistan (62)
Senegal (96)
Malawi (149)

## Appendix D. Expert Consultation Workshop on Artificial Intelligence as Public Goods

This expert consultation workshop was organized during the UN Open Source Week 2025 at UN Headquarters in New York on 17 June 2025, serving both as a launch of this research work and as a frame for initial research directions with stakeholders.

**Organizers:** Asian Development Bank (ADB), United Nations University Institute in Macau (UNU Macau), United Nations Office for Digital and Emerging Technologies (UN ODET).

**Number of participants:** 25

### WHY THIS WORKSHOP?

Artificial Intelligence (AI) is reshaping societies and economies worldwide. Yet its governance and accessibility remain contested. Applying Digital Public Goods (DPGs) principles to AI, creating AI as Digital Public Goods (AIDPGs), offers pathways towards transparency, inclusiveness, and global benefits.

This expert consultation workshop convened key knowledge experts to:
- Clarify the scope and definition of AIDPGs.

• Assess opportunities and risks.
• Explore global governance implications.
• Build consensus for future research and action.

**Key Opportunities Identified**
• Democratization of AI → wider access, autonomy, innovation.
• Stewardship & Reuse → open ecosystems enhance efficiency and adaptability.
• Global Challenges → potential for tackling large-scale social challenges, such as unemployment.
• Transparency & Trust → open models and shared governance mechanisms foster legitimacy.

**Key Risks & Barriers**
• “Rubber-Stamp” Endorsement → one-time approvals risk creating false trust.
• Governance Gaps → lack of clarity on contributors, committers, and accountability.
• North–South Divide → high barriers risk widening technological inequality.
• Open Washing → misuse of “public good” labels without real openness.
• Security & Misuse → political manipulation, bias, surveillance, and dual-use risks.

**Focal Issues Prioritized**
1. Transparency of sources and data – ensure accuracy, trust, and accountability.
2. Clear definitions – avoid confusion and “public goods washing”.
3. Governance and stewardship – iterative review, independent oversight, and sustainability.

**Emerging Insights**
• Definitions Matter: Without shared clarity, fragmentation threatens legitimacy.
• Trust is Foundational: Openness must be backed by responsible governance.
• Adaptive Governance: Static one-time approvals are insufficient; living, iterative models are needed.
• Sustainability Metrics: Long-term ecosystem health must be integrated into AIDPG Standard.
• Global Inclusiveness: Frameworks must balance ambition with practicality for the developing countries.

**Next Steps**
• Strengthen digital trust narratives linking openness → transparency → adoption.
• Co-develop governance models clarifying contributor/committer roles.
• Cluster use cases (e.g. health, privacy, epidemic control) to test governance models.
• Integrate sustainability metrics into standards.

**Conclusion**
The workshop confirmed both the urgency and opportunity in framing AI as Digital Public Goods. By embedding openness, inclusivity, and adaptive governance, AIDPGs can advance the SDGs while addressing risks of exclusion, bias, or misuse.

This consultation laid the foundation for future ADB–UNU Macau research and multi-stakeholder collaboration, with actionable priorities for building trustworthy and sustainable AI public goods for all.

## Appendix E. Ethical Review Board

All primary data collection, including interviews and surveys, was conducted in accordance with institutional ethical standards, receiving prior approval from the Joint Ethical Review Board (ERB) of the United Nations University (Ref. No: 202506/01).

## Appendix F. Artificial Intelligence Statement

**We are using an AI statement framework following the Contributor Roles Taxonomy (CRediT)[56] that is normally used to outline 'contributors' individual roles as authors of a research output (Weaver, 2024.)[57]**

*Artificial Tools:* Claude Opus 4.7 and Microsoft Copilot (UNU institutional instance); *Conceptualization:* no AI tools were used; *Methodology:* no AI tools were used; *Information Collection:* no AI tools were used; *Data Collection Method:* no AI tools were used; *Execution:* no AI tools were used; *Data curation:* no AI tools were used; *Data Analysis:* No AI tools were used; *Privacy and Security:* no identifiable data was shared with AI tools; *Interpretation:* no AI tools were used; *Visualization:* AI tools were used to design JSX data visualizations; *Writing - Review & Editing:* AI tools were used to refine the language, fix grammar issues and summarize long paragraphs; *Writing - Translation:* no AI tools were used; *Project Administration:* no AI tools were used.

**The authors assume full responsibility for the accuracy and integrity of the scientific content presented in this publication.**

# Footnotes

1 Oxford Insights. (2026). Government AI Readiness Index 2025. https://oxfordinsights.com/ai-readiness/government-ai-readiness-index-2025/; UNESCO. (2023). Evaluating national AI readiness with the Government AI Readiness Index. https://www.unesco.org/ethics-ai/en/articles/evaluating-national-ai-readiness-government-ai-readiness-index

2 https://legalinstruments.oecd.org/en/instruments/OECD-LEGAL-0449

3 UNESCO. (2021). Recommendation on the Ethics of Artificial Intelligence. https://unesdoc.unesco.org/ark:/48223/pf0000380455.locale=en

4 Digital Public Goods Alliance. (2025). The 2025 State of the Digital Public Goods Ecosystem. https://www.digitalpublicgoods.net/2025-DPG-Ecosystem-Report

5 Digital Public Goods Alliance GovStack Community of Practice. (2022). GovStack Definitions: Understanding the Relationship between Digital Public Infrastructure, Building Blocks & Digital Public Goods. https://www.digitalpublicgoods.net/DPI-DPG-BB-Definitions.pdf.

6 Digital Public Goods Alliance. (n.d.). Digital Public Goods Standard Governance. https://github.com/DPGAlliance/DPG-Standard/blob/master/governance.md

7 Principles for Responsible Investment. (n.d.). What is responsible investment? https://www.unpri.org/responsible-investment/intro-guides/what-is-responsible-investment.

10 Encyclopaedia Britannica. (n.d.). Free riding. https://www.britannica.com/topic/free-riding.

11 OECD. (n.d.). Generative AI. https://www.oecd.org/en/topics/generative-ai.html.

12 International Monetary Fund. (2021). What Are Global Public Goods? https://www.imf.org/en/publications/fandd/issues/2021/12/global-public-goods-chin-basics

13 United Nations ICT Task Force. (2003). Tools for Development: Using Information and Communication Technologies to Achieve the Millennium Development Goals. Working Paper. https://www.itu.int/net/wsis/stocktaking/docs/activities/1103056110/ICTMDGFinal.pdf

14 Jha, S. K., Gold, R., & Dubé, L. (2021). Modular Interorganizational Network Governance: A Conceptual Framework for Addressing Complex Social Problems. Sustainability, 13(18), 10292. https://doi.org/10.3390/su131810292

15 Oakland, W. H. (1987). Theory of public goods. In A. J. Auerbach & M. Feldstein (Eds.), Handbook of Public Economics (Vol. 2, pp. 485-535). Elsevier. https://doi.org/10.1016/S1573-4420(87)80004-6

16 Ibid.

17 Creative Commons. (n.d.). Open license. https://wiki.creativecommons.org/wiki/Open_license

18 Chuang, T.-R., Fan, R. C., Ho, M.-S., & Tyagi, K. (2022). Openness. Internet Policy Review, 11(1). https://doi.org/10.14763/2022.1.1643

19 Open Source Initiative. (n.d.). Frequently Answered Questions. https://opensource.org/faq

20 Open Source Initiative. (n.d.). The Open Source AI Definition 1.0. https://opensource.org/ai

21 Open Source United. (n.d.). https://opensource.un.org/en

22 Oakland, W. H. (1987). *Theory of public goods*. In A. J. Auerbach & M. Feldstein (Eds.), Handbook of Public Economics (Vol. 2, pp. 485-535). Elsevier. https://doi.org/10.1016/S1573-4420(87)80004-6.

23 OECD. (2024). *OECD Due Diligence Guidance for Responsible AI*. https://www.oecd.org/en/publications/oecd-due-diligence-guidance-for-responsible-ai_41671712-en.html

24 Open Source United. (n.d.). *United Nations Open Source Principles*. https://opensource.un.org/en/news/united-nations-open-source-principles

25 United Nations. (2019). The age of digital interdependence: Report of the UN Secretary-General's High-Level Panel on Digital Cooperation. https://www.un.org/en/pdfs/DigitalCooperation-report-for%20web.pdf

26 DPGA. (2021). 5 Year Strategy (2021-2026) – Promoting digital public goods to create a more equitable world. https://www.digitalpublicgoods.net/DPGA_Strategy_2021-2026.pdf

27 Digital Public Goods Alliance. (2025). The 2025 State of the Digital Public Goods Ecosystem.https://www.digitalpublicgoods.net/2025-DPG-Ecosystem-Report

28 United Nations Development Programme. (2022). 5 ways digital public goods are powering the Global Goals. https://www.undp.org/stories/5-ways-digital-public-goods-are-powering-global-goals

29 United Nations. (n.d.). Pact for the Future. https://www.un.org/en/summit-of-the-future/pact-for-the-future

30 United Nations. (n.d.). Open Source United. https://opensource.un.org/en

31 United Nations. (n.d.). UN Open Source Principles. https://opensource.un.org/en/news/united-nations-open-source-principles

32 https://www.un.org/en/content/digital-cooperation-roadmap/

33 https://evonomics.com/building-public-goods-21st-century/

34 https://digitallibrary.un.org/record/3865925?v=pdf

35 https://www.un.org/digital-emerging-technologies/global-digital-compact/intergovernmental-process

36 Sidhpurwala, H. (2025). The ethics of open and public AI: Balancing transparency and safety. Red Hat Blog. https://www.redhat.com/en/blog/ethics-open-and-public-ai-balancing-transparency-and-safety

37 https://opensource.org/ai

38 https://www.digitalpublicgoods.net/CoP-AI-Recommendations.pdf

39 Note: The authors' synthesis is based on a structured comparative document analysis of the CoP Discussion Paper and the subsequent CoP Recommendations documents.

40 The District Health Information Software 2 is a widely deployed open-source health management information system used by ministries of health in over 70 countries. https://dhis2.org/

41 The qualitative sample consisted of 21 knowledge experts and thought leaders representing policymaking (e.g. UN agencies, international organizations, government bodies), the private sector (start-ups and large tech companies), technical development (open source AI and DPG practitioners), civil society, and academia.

42 We captured insights from 37 experts across similar sectors, including representatives from prominent institutions at the forefront of the field, such as Creative Commons and MIT CSAIL.

43 The World Academy of Sciences. (n.d.). Developing Countries in the South. https://twas.org/developing-countries-south

44 World Bank. (n.d.). World Bank Country and Lending Groups. https://datahelpdesk.worldbank.org/knowledgebase/articles/906519-world-bank-country-and-lending-groups

45 Oxford Insights. (2025). Government AI Readiness Index 2025. https://oxfordinsights.com/ai-readiness/government-ai-readiness-index-2025

46 This is aligned with the International Dialogue on AI Safety (IDAIS) “Consensus Statement On AI Safety As A Global Public Good” delivered in Venice in 2024. https://idais.ai/dialogue/idais-venice

47 A position reaffirmed by the DPGA–UNICEF Community of Practice (Gimpel, 2024).

48 Meta Platforms, Inc. (2024). Llama 3.1 Community License Agreement. https://www.llama.com/llama3_1/license

49 https://brasil.mapbiomas.org/en/

50 https://www.digitalpublicgoods.net/r/modular-open-source-identity-platform

51 https://www.un.org/digital-emerging-technologies/

52 United Nations Development Programme. (2021). UNDP joins the Digital Public Goods Alliance to accelerate inclusive digital transformation. https://www.undp.org/news/undp-joins-digital-public-goods-alliance-accelerate-inclusive-digital-transformation

53 https://aiforgood.itu.int/reports_publications/ai-for-good-impact-report-2025-2nd-edition/

54 https://www.adb.org/news/events/aixadb-2025

55 https://www.limesurvey.org/

56 https://credit.niso.org/

57 https://journals.sagepub.com/doi/10.1177/18758789261435716